\documentstyle[11pt,aaspp4]{article}
\begin{document}
%
%
\def\etal{{\it et al.}\ }
\def\abs#1{\left| #1 \right|}
\def\EE#1{\times 10^{#1}}
\def\gcm{\rm ~g~cm^{-3}}
\def\kms{\rm ~km~s^{-1}}
\def\ergs{\rm ~erg~s^{-1}}
\def\isotope#1#2{\hbox{${}^{#1}\rm#2$}}
\def\wl{~\lambda~}
\def\wll{~\lambda~\lambda~}
\def\HI{{\rm H\,I}}
\def\HII{{\rm H\,II}}
\def\HeI{{\rm He\,I}}
\def\HeII{{\rm He\,II}}
\def\HeIII{{\rm He\,III}}
\def\CI{{\rm C\,I}}
\def\CII{{\rm C\,II}}
\def\CIII{{\rm C\,III}}
\def\CIV{{\rm C\,IV}}
\def\NI{{\rm N\,I}}
\def\NII{{\rm N\,II}}
\def\NIII{{\rm N\,III}}
\def\NIV{{\rm N\,IV}}
\def\NV{{\rm N\,V}}
\def\NVI{{\rm N\,VI}}
\def\NVII{{\rm N\,VII}}
\def\OI{{\rm O\,I}}
\def\OII{{\rm O\,II}}
\def\OIII{{\rm O\,III}}
\def\OIV{{\rm O\,IV}}
\def\OV{{\rm O\,V}}
\def\OVI{{\rm O\,VI}}
\def\CaI{{\rm Ca\,I}}
\def\CaII{{\rm Ca\,II}}
\def\NeI{{\rm Ne\,I}}
\def\NeII{{\rm Ne\,II}}
\def\NaI{{\rm Na\,I}}
\def\NaII{{\rm Na\,II}}
\def\NiI{{\rm Ni\,I}}
\def\NiII{{\rm Ni\,II}}
\def\CaII{{\rm Ca\,II}}
\def\FeI{{\rm Fe\,I}}
\def\FeII{{\rm Fe\,II}}
\def\FeIII{{\rm Fe\,III}}
\def\FeIV{{\rm Fe\,IV}}
\def\FeV{{\rm Fe\,V}}
\def\FeVII{{\rm Fe\,VII}}
\def\CoII{{\rm Co\,II}}
\def\CoIII{{\rm Co\,III}}
\def\ArI{{\rm Ar\,I}}
\def\ArII{{\rm Ar\,II}}
\def\MgI{{\rm Mg\,I}}
\def\MgII{{\rm Mg\,II}}
\def\MgIII{{\rm Mg\,III}}
\def\SiI{{\rm Si\,I}}
\def\SiII{{\rm Si\,II}}
\def\SiIII{{\rm Si\,III}}
\def\SiIV{{\rm Si\,IV}}
\def\SI{{\rm S\,I}}
\def\SII{{\rm S\,II}}
\def\SIII{{\rm S\,III}}
\def\FeI{{\rm Fe\,I}}
\def\FeII{{\rm Fe\,II}}
\def\FeIII{{\rm Fe\,III}}
\def\FeIIV{{\rm Fe\,IV}}
\def\kI{{\rm k\,I}}
\def\kII{{\rm k\,II}}
\def\La{{\rm Ly}\alpha}
\def\Ha{{\rm H}\alpha}
\def\Hb{{\rm H}\beta}
\def\Hg{{\rm H}\gamma}
\def\He{{\rm H}\epsilon}
\def\Paa{{\rm Pa}\alpha}
\def\Pag{{\rm Pa}\gamma}
\def\Bra{{\rm Br}\alpha}
\def\Brg{{\rm Br}\gamma}
\def\Lya{{\rm Ly}\alpha}
\def\Msun{{~\rm M}_\odot}
\def\Msunyr{M_\odot~yr^{-1}}
\def\tyr{t_{\rm yr}}
%
%
\title{Late Spectral Evolution of SN 1987A:\\
I. Temperature and Ionization}
\author{Cecilia Kozma and Claes Fransson}
\affil{Stockholm Observatory,
S-133$\,$36 Saltsj\"obaden, Sweden}
\authoremail{cecilia@astro.su.se, claes@astro.su.se}
\begin{abstract}
The temperature and ionization of SN 1987A  is modeled between 200 and
2000 days in its nebular phase, using a
time-dependent model. We include all important elements, as well as
the primary composition zones in the supernova.
The energy input is provided by radioactive decay of \isotope{56}{Co},
\isotope{57}{Co}, and \isotope{44}{Ti}. 
The thermalization of the resulting gamma-rays and positrons is
calculated by
solving the Spencer-Fano equation. Both the ionization and the
individual level
populations are calculated time-dependently. Adiabatic cooling is
included in
the energy equation. Charge transfer is important for determining the
ionization, and is included with available and estimated rates.
Full, multilevel
atoms are used for the observationally important ions. 
As input models to the calculations we use
explosion models
for SN 1987A calculated by Woosley \etal and Nomoto \etal

The most important result in this paper refers to the evolution of the
temperature and ionization  of the various abundance zones. 
The metal-rich core undergoes a thermal instability,
often referred to as the IR-catastrophe,
at $600 - 1000$ days.
The hydrogen-rich zones evolve
adiabatically after $500 - 800$ days, while in the helium region both
adiabatic cooling and 
line cooling are of equal importance after $\sim 1000$ days. Freeze-out of the
recombination 
is important in the hydrogen and helium zones. Concomitant with the
IR-catastrophe, 
the bulk of the emission shifts from optical and near-IR lines to the
mid- and far-IR. After the IR-catastrophe, the cooling is mainly due
to far-IR lines and adiabatic expansion.

Dust cooling is likely to be 
important in the zones where dust forms.
We find that the dust condensation temperatures occur later than $\sim
500$ days in the oxygen-rich zones, and the most favorable zone for
dust condensation is the iron core. 
The uncertainties introduced by the, in some cases,
unknown charge transfer
rates are discussed. Especially for ions of low abundances differences
can be substantial.

\end{abstract}


\section{INTRODUCTION}
The observations of SN 1987A represent a unique set of
data. Especially at late epochs high
quality spectra exist at epochs when other supernovae are too faint
to be observed even photometrically. This is particularly true for
the IR region (e.g., Chevalier 1992, McCray 1993). The analysis of
these observations are, however, not trivial and represent a
considerably more complex situation compared to that of a
planetary nebula, or an H II region. Complications include large optical
depths of even forbidden lines, high velocities and consequent severe line
blending, non-thermal excitation and ionization, non-uniform
abundances, molecule and dust formation, as well as time-dependent
effects. For reviews of the physics see McCray (1993) and  Fransson
(1994). In this paper and in Kozma \& Fransson (1997) (Paper II)  we concentrate on the spectral modeling in the
nebular phase when the core region is transparent in at least the
optical and IR regions. In a subsequent paper  we will
discuss the broad-band photometry, including the bolometric light curve.

Previous discussions of the late, nebular spectrum can be divided into
attempts of 
self-consistent modeling of the spectrum, or analyses of the lines of
particular elements or molecules. Models in the first category are
hampered in many cases by lack of atomic data, as well as the general
complexity of the problem. Examples of this type of models include some
early predictions by Fransson
\& Chevalier (1987) and Colgan \& Hollenbach (1988), and modeling of 
observed spectra by Swartz, Harkness, \& Wheeler  (1989). Applications to core collapse supernovae in general, including
Type Ib's and Type Ic's can be found in Fransson \& Chevalier (1989)
and Swartz (1994). The 'diagnostic'  approach has
been  discussed in a series of papers by Li
\& McCray (1992, 1993, 1995), Li, McCray \& Sunyaev (1993), Xu \etal
(1992),  Chugai (1994), and by Kozma
\& Fransson (1992, hereafter KF92). While the  problem in
this case is simplified to an easily controllable problem, it is often
limited by the lack of knowledge of temperature, ionization and radiation
field. 
As will be apparent in this paper and in Paper II, the limitation to one specific
chemical composition, and in most cases a single element or ion, is
especially severe.

Predictions are one of the best methods for judging the success
of a complex model. It is therefore of some interest to discuss the
paper by Fransson \& Chevalier (1987), where immediately after the
explosion predictions for the
spectral evolution were made. There are several limitations to that paper. The most
serious were the neglect of the hydrogen zone, and the simplified
treatment of the iron emission. The model was therefore only applicable to the
helium, oxygen and silicon -- sulphur regions. Even with these
limitations there were several interesting results.
An important
prediction in this paper 
was the dramatic drop in temperature at $\sim 600 - 700$ days in
the metal-rich regions, and consequent transition from optical to
far-IR radiation. This is commonly referred to as the IR-catastrophe (e.g., Axelrod 1980, Fransson 1994). The optical to far-IR
transition 
is one of the characteristic features in the evolution of SN
1987A, and is clearly reflected if the light
curves of the lines are plotted as a fraction of the total bolometric
luminosity (e.g., Menzies 1991).
Often this has been attributed to dust thermalization, which is
certainly important, but independent of this, just the
normal cooling in the metal-rich core will have this effect.
The drop in temperature is also most likely a requirement for molecule
and dust formation, and was in fact anticipated by
Fransson \& Chevalier.
The concomitant
drop in the optical lines and the dust formation, as seen in the
line profiles (Lucy \etal 1989, 1991) and mid-IR emission (e.g., Wooden
\etal 1993), are therefore
most likely connected.

The predictions by Fransson \& Chevalier have been directly compared
to the observations of the light curves of the strongest lines,  by
Menzies (1991), Spyromilio \etal (1991), and
Meikle \etal (1993). While there are quantitative
discrepancies, it was found that the general
shape of e.g. the strong [Ca II] $\wll 7291, 7324$ and [O I] $\wll 6300,
6364$ light curves agreed well qualitatively with the
predictions. Note that most of the [Ca II] emission in this model
comes from the oxygen-rich zone, and not from the calcium-rich Si -- S
zone, showing that a small amount of an effective cooler, like Ca II,
can dominate the emission from a zone with a large gamma-ray
deposition. In Paper II we will see that this is also the case in
these models, although it is the hydrogen-rich zones, rather than the
oxygen-rich wich dominate the emission, in agreement with Li \& McCray
(1993). 

 Another
interesting result in the paper by Fransson \& Chevalier was the
non-uniform temperature and 
ionization in the different composition zones, with the metal-rich
regions having considerably lower temperature. This illustrates the
necessity to take all important abundance components into account, as
well as the importance of distinguishing between macroscopic mixing of
blobs of different compositions, compared to  microscopic mixing of
the gas. 

An important aspect of the late evolution of the spectrum and light
curve is the departure from (quasi-) steady state. As density
decreases  $\propto t^{-3}$,
recombination and cooling time scales 
increase. Eventually, as these become longer than the radioactive decay
time scale, $\tau$, the spectrum and light curve have to be calculated
time-dependently. In addition, when the radiative  cooling time
becomes longer than the expansion time scale, adiabatic cooling must be
included. These effects were discussed in Fransson \& Kozma (1993),
with particular emphasis on the consequences for the bolometric light
curve. Implicit in this calculation was, however, a complete
calculation of the spectrum as a function of time. Some results of this
were reported in Fransson, Houck \&
Kozma (1996), and it was shown  especially that
the [O I] $\wll 6300, 6364$ lines provided compelling evidence for the
IR-catastrophe. In addition, the importance of non-thermal excitation
for the spectrum at late time were stressed. These points are
discussed further in Paper II. 

In these two papers we continue our attempt to model the evolution of the
line emission from 
SN 1987A as realistically as possible. 
Consequently, we calculate the
temperature and ionization self-consistently, using a time-dependent
formalism. Non-thermal processes are included with the method in
KF92. In addition, we include all
important abundance zones in 
the ejecta with appropriate filling factors. The important ions are
modeled as multilevel atoms, using the best available atomic data.
Even with these ingredients there are a number of deficiencies in our
modeling. Probably the most serious omission is that  we do not
include the resonance line blanketing, especially 
important in the UV. Further, the
molecular chemistry and cooling are not included. Although
we have tried to include
the best atomic data
available, we are still handicapped by a lack of reliable data, especially with regard to charge transfer
reactions and photoionization  of the iron group elements.
In Paper II we argue that these omissions,
although important for specific lines, should not influence our general
conclusions severely. We  hope to improve upon these shortcomings in
the future.

In this paper we discuss our model and the temperature and ionization
evolution in detail. In Paper II we concentrate on the line emission,
as well as on the line profiles.
In section ~\ref{sec-model} we discuss our basic model, assumptions, and
physical processes included, as well as the explosion models we use.
Section ~\ref{sec-results}  contains a detailed discussion of our
results with regard to the temperature and ionization, while in
section ~\ref{sec-ct2} we discuss the sensitivity of our
results to especially the UV-radiation field and the poorly known
charge transfer rates. In
section ~\ref{sec-dustcool} we
discuss the dust formation and the effects of dust cooling. In section
\ref{sec-conclusion} 
our main results are for convenience summarized.
In the appendix we have collected the references for the atomic data.

\section{THE MODEL}
\label{sec-model}
In these papers we focus on  the emission from SN 1987A at times later 
than $\sim$ 200 days, when
the supernova has entered the nebular phase. 
As the  matter is expanding and becoming more transparent,  
it is possible to extract information 
on the nucleosynthesis, which has taken
place in the progenitor and in the explosion, as well as on the
hydrodynamics of the explosion.

Our model  calculates the temperature, ionization and level
populations of the
different regions in the supernova envelope. The first time step is
calculated in steady state, which is a good approximation at 200 days.
The following time steps are calculated time-dependently. Knowing the
time evolution of the physical properties of the supernova matter we
calculate the bolometric luminosity, as well as luminosities for
$\sim 5900$ lines. As input, the calculations require an
explosion model, as well as the radial structure of
the ejecta with abundances and masses as a function of velocity.
Our model is spherical symmetric, with zones of different
compositions and thicknesses. Mixing in velocity is taken into account
in an approximate way by mixing shells from different abundance zones
in velocity.

We have the option to switch between either doing the calculations in
steady state or time-dependently. 
The steady state calculations are done primarily in order to
check that energy is conserved at every time step. We find that energy is
conserved to within $\sim$ 1 \% at all times.
By comparing the steady state and time-dependent calculations it is
possible to determine the level of freeze-out.
The temperature, ionization, and level populations of the
multi-level atoms are calculated time-dependently.

Below, we first discuss the physical processes
included in our model, and then the explosion models used in
our calculations. 

\subsection{Physical Processes}
\subsubsection{Temperature Calculation}
\label{sec-tempcalc}
The temperature, $T$, of a region is determined by balancing the cooling and
heating mechanisms. 
In steady state the total cooling ($\Lambda n n_e$) at a certain
point is just balanced by the total heating ($H$),
\begin{equation} 
\Lambda(T) n n_e = H .
\end{equation}
From this condition it is possible to determine the temperature.
In the time-dependent case the temperature change  can
be expressed as 
\begin{equation}
\label{eq:dTdt}
\frac{dT}{dt} = \frac{2}{3k(1 + x_e)n} (H - \Lambda(T) n n_e) - \frac{2
T}{t} - \frac{T}{1 + x_e} \frac{dx_e}{dt} 
\end{equation}
where $n$ is the number density of ions and atoms, $n_e$ the number
density of free electrons, and $x_e = n_e/n$.
The second term in this equation describes adiabatic cooling. If
this term dominates, $T \propto t^{-2}$.

Most of the cooling is by collisional excitation of 
bound-bound transitions. 
The total line cooling is a sum over all the transitions $i, j$,
\begin{equation}
\label{eq:lincool}
L_{\rm lines} = \sum_{i,j} (n_i n_e C_{ij}(T) - n_j n_e C_{ji}(T)) h\nu_{ij}
\hspace{1.0cm},
\end{equation}
where $C_{ij}(T)$ and $C_{ji}(T)$ 
are the collisional excitation, and
de-excitation rates, respectively (Osterbrock 1989), $n_i$
is the number densities of ions in level $i$, and $h\nu_{ij}$ is the
photon energy. 
Line cooling is the dominant cooling process, but we also include
cooling due to recombinations and free-free cooling.
We approximate  recombination cooling with
\begin{equation} 
L_{\rm rec}  \approx  \sum_{k} n_{k+1} n_e \alpha_k kT
\hspace{1.0cm}, 
\end{equation}
where the summation is over all ions $k$, and $\alpha_k$ refers
to the recombination from $k+1$ to $k$. 

Free-free cooling is given by 
\begin{eqnarray}
L_{\rm ff} & = & \sum_k 1.426 \EE{-27} T^{1/2} Z_k^2 n_e n_k <g_{\rm ff}>.   
\end{eqnarray}
The summation is over the ions $k$, $Z_k$ is the charge of ion $k$,
and $<g_{\rm ff}>$ is the mean gaunt factor for free-free emission. 

Cooling, both due to lines, recombinations, and free-free emission, is
in the low density,
optically thin limit, proportional to
the electron fraction. We therefore 
define a cooling function, $\Lambda (T)$, so that the total cooling
equals $\Lambda(T) n n_e = L_{\rm lines} + L_{\rm rec} + L_{\rm ff}$,
\begin{eqnarray}
\Lambda(T) & = & \sum_{k}[\sum_{ij}(x_i C_{ij}(T) - x_j C_{ji}(T))
h\nu_{ij} +  \cr
& & {3 \over 2} x_{k+1} \alpha_k kT + 
1.426 \EE{-27} T^{1/2} Z_k^2 x_k
<g_{\rm ff}>] 
\hspace{1.0cm},
\end{eqnarray}
where $x_i$ is the fraction of ions in level $i$, and the summation
over $k$ represents all ions. $\Lambda$ depends
sensitively on the electron temperature, $T$.

The heating in the ejecta is dominated by  non-thermal heating,
$H_{\gamma}$, 
which is independent of the electron temperature. This process is
discussed in detail in KF92.
We  express the non-thermal heating as
\begin{equation}
H_{\gamma} = 4\pi J_{\gamma} \sigma_{\gamma} \eta_{\rm H} n
\hspace{1.0cm},
\end{equation}
where $4\pi J_{\gamma} \sigma_{\gamma}$ is the deposited non-thermal
energy per particle. This quantity is discussed in more detail
below. $\eta_{\rm H}$ is 
the fraction of the deposited energy going into heating the free,
thermal electrons. $\eta_{\rm H}$ depends  on the electron
fraction, and increases with increasing $x_e$ (KF92).  

Photoelectric heating from the ground states is included for all
ions $k$, and  from excited levels $i$, where there is atomic data
available. 
The total photoelectric heating can be expressed,
\begin{equation}
H_{\rm P} = \sum_k \sum_i \hspace{0.3cm} n_{k,i} \int_{\nu_{k,i}}^{\infty}
4 \pi \frac{\sigma_{\nu} J_{\nu} (h\nu - \chi_{k,i})}{h\nu} d\nu
\hspace{1.0cm},
\end{equation}
where $h \nu_{k,i}$ is the ionization energy, $\sigma_{\nu}$ is the
photoionization cross section 
from level $i$ in ion $k$, $J_{\nu}$ 
 is the mean radiation intensity at $\nu$, and $\chi_{k,i}$
is the ionization potential.

The total heating is therefore
\begin{equation}
H = H_{\gamma} + H_{\rm P} \hspace{1.0cm}.
\end{equation}

\subsubsection{Ionization Balance}
In our calculations we include H I-II, He I-III, C I-III, N I-II, O
I-III, Ne I-II, 
Na I-II, Mg I-III, Si I-III, S I-III, Ar I-II, Ca I-III, Fe
I-V, Co II, and Ni I-II.

For an estimated value of the electron fraction, $x_e$,
the ionization balance for an element with only two ionization stages, $k$
and $k+1$  is solved from 
\begin{eqnarray}
\label{eq:ionbal}
\frac{ dn_k}{dt} + \frac{ 3 n_k}{t} & = & - (\Gamma_{k} + P_{k}) n_{k} 
- n_k \sum_l n_l \xi^{\rm CT}_{k,l} + \cr
 & & \alpha_{k} n_e n_{k+1} 
+ n_{k+1} \sum_l n_l \xi^{\rm CT}_{k+1,l} .
\end{eqnarray}
For  elements with three or more ionization stages equation
(\ref{eq:ionbal}) is 
generalized to include processes both below and above the relevant
ionization stage. 
 Here, $\Gamma_{k}$ is the non-thermal
ionization rate, 
\begin{equation}
\label{eq:gamion}
\Gamma_{k} = 4\pi \frac{J_{\gamma} \sigma_{\gamma}}{\chi_{eff,k}} 
\hspace{1.0cm},
\end{equation}
where $\chi_{eff,k}$ is the effective ionization potential, defined as
$\chi_{eff,k} = \chi_k x_k / \eta_k $ (see KF92), $\chi_k$ being the
ionization potential, $x_k$ the number fraction, and $\eta_k$ the
fraction of the deposited non-thermal energy going into ionization of ion
$k$. 
$P_{k}$ is the photoionization rate, 
\begin{equation}
P_k = 4\pi \int_{\nu_0}^{\infty} \frac{\sigma_{k,\nu}J_{\nu}}{h\nu} d\nu
\hspace{1.0cm},
\end{equation}
while $\alpha_{k}$ is
the recombination coefficient from $k+1$ to $k$. In our
calculations $P_{k}$ and $\alpha_{k}$ include photoionizations
from, and recombinations to excited levels in ion k. 

In many cases charge transfer reactions with ions $l$ may
contribute, either as an ionization, $n_k n_l \xi^{\rm CT}_{k,l}$, 
or recombination, $n_{k+1} n_l \xi^{\rm CT}_{k+1,l}$, process
(Fransson \& Chevalier 1989, Swartz 1994). Excited levels of ion $k$ may
also be involved.

Finally, the number conservation equation is
\begin{equation}
\label{eq:numcons}
\sum_k n_k = n_{\rm element}.
\end{equation}
Equations (\ref{eq:ionbal}) and (\ref{eq:numcons}) are solved using the Newton-Raphson
method. For stability and 
efficiency,
 equations (\ref{eq:ionbal}) and (\ref{eq:numcons}) are solved
implicitly with backward differencing,  
\begin{equation}
\frac{n_k^{n+1} - n_k^{n}}{\Delta t} = f(n_k^{n+1})
\end{equation}
where $n_k^{n+1}$ refers to the new value and $n_k^{n}$ to the old
value. The function $f(n_k^{n+1})$
on the right hand side includes all ionization and recombination
processes in equation (\ref{eq:ionbal}). 
Even though the implicit treatment requires iteration, it makes
the overall problem more stable, and allows considerably longer time steps. 
The time step varies between the different zones, depending on the
cooling and recombination time scales. The hydrogen and helium-rich
regions therefore require  smaller number of time steps, compared to
the metal-rich regions, where these time scales are short. For the same
reason, the late stages allow longer time steps, compared to the
first epochs. As an example, the iron zones require $\sim 15000$ time
steps from 200 -- 2000 days, while the corresponding numbers are $\sim
4000$ for the oxygen zones, and $\sim 1000$ for the hydrogen zones. 

\subsubsection{Level Populations}
With the
abundance of each ionization stage known, the level
populations are calculated for a number of ions. For the following 
multilevel atoms 
we calculate the level populations time-dependently :
\HI~ (30 levels with $n \leq 20$), \HeI~ (16 levels), \OI~ (13 levels), \CaII~ (6 levels),
\FeI~ (121 levels),  \FeII~ (191 levels), \FeIII~ (110 levels), and 
\FeIV~ (43 levels). 
Lines from other elements are solved, either as two-level
atoms, or, in the case of forbidden lines, as 4-6 level systems (see
appendix).  

The rate equations for the level populations are given by
\begin{eqnarray}
\frac{dn_{k,i}}{dt} + \frac{3 n_{k,i}}{t}&  = &  - n_{k,i} (P_{k,i} +
\Gamma_{k,i} + \alpha_{i,k-1} n_e)  \cr
 & & \cr
 & & - n_{k,i}  \sum_{j<i}( A_{ij} \beta_{ij} +
C_{ij} n_e + P_{ij}) \cr
 & & \cr
 & &  + \sum_{j<i} n_{k,j} (C_{ji} n_e +\Gamma_{ji} + P_{ji}) \cr
 & & \cr
 & &  - n_{k,i} \sum_{j>i} (C_{ij} n_e + \Gamma_{ij} + P_{ij})   \cr
 & & \cr
 & & + \sum_{j>i} n_{k,j} (A_{ji} \beta_{ji} + C_{ji} n_e + P_{ji})   \cr
 & & \cr
 & & + n_{k+1} \alpha_{k,i} n_e + n_{k-1} (P_{i,k-1} + \Gamma_{i,k-1}) \cr
 & & \cr
 & & - n_{k,i} \sum_l n_l \xi^{\rm CT}_{k,l} + n_{k+1} \sum_l n_l 
\xi^{\rm CT}_{k+1,l}  \cr 
 & & \cr
 & & + n_{k-1} \sum_l n_l \xi^{\rm CT}_{k-1,l}
\end{eqnarray}
where the index $i$ refers to the actual level, and $j$ to other levels
within the ion, $k$ to the ionization stage of the element, and $l$ to any
other element.
$P_{k,i}$ and $\Gamma_{k,i}$ are the photoionization
rate and non-thermal ionization rate, from ion $k$, level $i$.
$\alpha_{i,k-1}$ is the recombination coefficient from ion $k$, and
level $i$ to ion $k-1$.  
Downward and upward radiative, and collisional rates,  
$ A_{ij} \beta_{ij}$ and $C_{ij} n_e$, are included. Here $\beta_{ij}$
is the escape probability, which includes both the Sobolev escape
probability and a continuum destruction probability. This is discussed
in more detail in sections~\ref{sec-lintran} and ~\ref{sec-contdest}.

Similar to the non-thermal ionization rate (eq. [\ref{eq:gamion}]) the
non-thermal 
excitation rate $\Gamma_{ij}$  from the lower level $i$
to the upper level $j$ is given by
\begin{equation}
\Gamma_{ij} = \frac{4\pi J_{\gamma} \sigma_{\gamma}}{\chi_{ij} x_i /
\eta_{ij}}  
\hspace{1.0cm},
\label{eq:gammion}
\end{equation}
where $\chi_{ij}$ is the excitation energy, $x_i$ the number fraction in level
$i$, and $\eta_{ij}$ is the fraction of the deposited energy going into
excitation of the transition $ij$.
For those elements
where non-thermal excitations are included, only excitations from the
ground state, $i=1$, are calculated except for O I, 
\FeI~ and \FeII.  For these, non-thermal excitations from
the individual levels of the ground state multiplet are included.

$P_{ij}$ and $P_{ji}$ are the photoexcitation/deexcitation rates
between levels $i$ and $j$. 
The Einstein relations and the Sobolev
assumption (e.g., Tielens \& Hollenbach 1985) result in 
\begin{equation}
P_{ji} = \frac{c^2}{2h\nu^3} A_{ji} \beta_{ji} J_{ji}
\end{equation}
and
\begin{equation}
P_{ij} = \frac{g_j}{g_i} P_{ji}.
\end{equation}

In accordance to our neglect of line blanketing, we only include the
continuum contribution to $J_{ji}$ (see section~\ref{sec-phie}).
Further, $\alpha_{k,i}$ is the recombination coefficient from ion
$k+1$ to an individual level $i$ of  ion $k$.
For ions
once or more ionized, photoionization and non-thermal ionization 
from ion $k-1$ are included. We assume that these ionizations end up
in the first level, $i=1$, $P_{i=1,k-1}$ and $ \Gamma_{i=1,k-1}$, with
the exception of O II. For oxygen we include non-thermal ionization
from O I to the excited states $^2D_o$ and $^2P_o$, as well as to the
ground state of O II.   

Charge transfer reactions with other elements may either act as
an ionization or recombination process. We assume that charge
transfer reactions occur from their ground states,
$i=1$, with the exception for He I. Swartz (1994) gives  rates for
 charge transfer involving excited levels in He I. 
The term $n_k n_l \xi^{\rm CT}_{k,l}$ represents the charge transfer rate for the
reaction 
$A^k + B^l \rightarrow A^{k\pm1} + B^{l\mp1}$, and
$n_{k\pm1} n_l \xi^{\rm CT}_{k\pm1,l}$ the charge transfer rate for the reaction
$A^{k\pm1} + B^l \rightarrow A^k + B^{k\pm1}$. 
Charge transfer reactions are further discussed in section~\ref{sec-CT}.

Together with the particle conservation equation
\begin{equation}
n_k = \sum_i n_{k,i},
\end{equation}
the rate equations are again
solved implicitly, using Newton-Raphson iteration.

\subsubsection{Non-Thermal Deposition}
The energy source for the ejecta at late times is well established 
to be the decay of radioactive isotopes created in the explosion.
The dominant isotope synthesized is \isotope{56}{Ni}. \isotope{56}{Ni}
rapidly decays to \isotope{56}{Co} on a time scale of $\tau_{\rm decay}$ = 8.8
days. \isotope{56}{Co} then decays to \isotope{56}{Fe} on a longer
time scale of 111.26 days. In the decay of \isotope{56}{Co} 96.5 \% of
the energy is emitted in the form of gamma-rays while the remaining
3.5 \% of the 
energy is emitted as positrons. Other important radioactive isotopes
created in 
the explosion are \isotope{57}{Ni} and
\isotope{44}{Ti}. \isotope{57}{Ni} decays  rapidly to \isotope{57}{Co},
and \isotope{57}{Co} to \isotope{57}{Fe} with
$\tau_{\rm decay}$ = 391.2 days. Decays of  \isotope{57}{Co} only
result in gamma-rays, no positrons. \isotope{44}{Ti} decays first
to \isotope{44}{Sc} 
on a time scale of 83.7 years followed by a rapid decay to \isotope{44}{Ca}.
In the decay of \isotope{44}{Sc} positrons are also emitted.

The gamma-ray luminosity due to the decay of \isotope{56}{Co} is
\begin{equation}
L_{\gamma} = 1.26 \EE{42} \frac{M(\isotope{56}{Ni})}{0.1 \Msun}
e^{-t/111.26^d} \hspace{1.0cm} {\rm erg ~ s^{-1}}   \label{eq:lum56}
\end{equation}
of \isotope{57}{Co},
\begin{equation}
L_{\gamma} = 1.37 \EE{41} M(\isotope{57}{Ni}) e^{-t/391.2^d}
\hspace{1.0cm} {\rm erg ~ s^{-1}}, \label{eq:lum57}
\end{equation}
and of \isotope{44}{Ti},
\begin{equation}
L_{\gamma} = 4.1 \EE{36} \frac{M(\isotope{44}{Ti})}{10^{-4} \Msun}
e^{-t/83.7^{yr}} \hspace{1.0cm} {\rm erg ~ s^{-1}}.  \label{eq:lum44}
\end{equation}
These expressions include the gamma-rays due to positron annihilation.

Positrons are assumed to be deposited locally in the iron-rich
regions. This is supported even at 7.8 years by the analysis of the
spectrum of SN 1987A by Chugai
\etal (1997).
The luminosity due to the energy input from  positrons from
\isotope{56}{Co} decay is
\begin{equation} 
L_{e^+} = 3.34 \EE{40} \frac{M(\isotope{56}{Ni})}{0.075 \Msun}
e^{-t/111.26^d}  \hspace{1.0cm} {\rm erg ~ s^{-1}}
\end{equation}
and from \isotope{44}{Ti},
\begin{equation} 
L_{e^+} = 1.3 \EE{36} \frac{M(\isotope{44}{Ti})}{10^{-4} \Msun}
e^{-t/83.7^{yr}}  \hspace{1.0cm} {\rm erg ~ s^{-1}}.
\end{equation}

The life time of $\isotope{44}{Ti}$ is uncertain, with values ranging
from 66.9 -- 96.1 years (see Timmes \etal 1996). Here we use the
latest value of 
$83.7 \pm 15$ years (Meissner \etal 1995).
In our calculations we include 0.07 $\Msun$ of \isotope{56}{Ni}, from
the bolometric light curve, $3.3
\EE{-3} \Msun$ of \isotope{57}{Ni}, which corresponds to a 
\isotope{57}{Ni}/\isotope{56}{Ni} ratio equal to 2 times the solar 
\isotope{57}{Fe}/\isotope{56}{Fe} ratio, and $10^{-4} \Msun$ of
\isotope{44}{Ti} 
(Woosley, Pinto \& Weaver 1988, Kumagai \etal 1991, Woosley \&
Hoffman 1991).
These masses correspond to our standard case in Fransson \& Kozma (1993).
With these masses, \isotope{56}{Co} dominates the energy input up to
$\sim$ 1200 days. Thereafter, \isotope{57}{Co} takes over up to $\sim$
1900 days, after which \isotope{44}{Ti}  dominates.  ${^{22}}$Na with
a decay time of 3.75 years never contributes more than 10 \% before
2000 days (Woosley, Pinto \& Hartman 1989).

Thermalization of the gamma-rays and positrons is discussed in detail
in  KF92 and Li, McCray, \& Sunyaev (1993). Using the
Spencer-Fano formalism (e.g., Spencer \& Fano 1954, Douthat 1975a,b, Xu
1989) we calculate the fractions of the deposited
energy  going into heating, ionization, and excitation,
respectively. These fractions depend on the composition of the
gas, the state of ionization and the electron fraction.
The Spencer-Fano calculations are repeated for each composition zone,
when the electron fraction has changed by 
more than 5 \%, or by more than 0.01 in absolute value.

The gamma-ray intensity depends on the distribution of the
radioactive elements. This is uncertain, and depends on the
degree of mixing in the gas (see section ~\ref{sec-explosionmodel}). 
We distribute  the radioactive source  within the core, proportional to
the iron mass. 
The ionization, excitation, and heating depend on  $4\pi
J_{\gamma} \sigma_{\gamma}$ (${\rm erg ~ s^{-1} ~ particle^{-1}}$),
the total gamma-ray energy 
deposited per second per particle at a  certain radius in the
supernova ejecta.
With  particles we refer to ions and atoms ({\em not} electrons).
$J_{\gamma}(r)$
is the gamma-ray mean intensity
at radius $r$, $\sigma_{\gamma} (r)$ (${\rm cm^{2} ~ particle^{-1}}$)
is the energy-averaged cross 
section for gamma-ray absorption at radius $r$ for an ion. 
For the gamma-rays resulting from the decay of \isotope{56}{Co},
Fransson \& Chevalier (1989) find  $\kappa_{\gamma} = 0.06 ~ 
(Z_i/A_i) ~ {\rm cm^2 ~g^{-1}}$, where $Z_i$ and $A_i$ is the nuclear
charge and the atomic weight, respectively, for ion $i$.
For hydrogen-rich regions $Z_i/A_i \approx 1$, while in the other
regions, dominated by helium or heavier elements, $Z_i/A_i \approx
0.5$. Therefore, we use  $\kappa_{\gamma}(\isotope{56}{Co}) =
0.06 ~ {\rm cm^2 ~g^{-1}}$
in the hydrogen-rich regions, and $\kappa_{\gamma}(\isotope{56}{Co}) =
0.03 ~ {\rm cm^2 ~g^{-1}}$
in the remaining regions. The average cross section for a particular
composition is  then calculated from
$\sigma_{\gamma}(\isotope{56}{Co}) = A_{\rm mean} ~ m_p ~
\kappa_{\gamma}(\isotope{56}{Co})$, where $A_{\rm mean}$ is the mean
atomic weight in the region. 

The mean energy of the gamma-rays emitted in  the
\isotope{57}{Co}-decay is lower than in the decay of
\isotope{56}{Co}. Woosley \etal (1989) find that  
$\kappa_{\gamma}(\isotope{57}{Co}) \approx 2.4 ~
\kappa_{\gamma}(\isotope{56}{Co})$, and 
also an effective opacity for  the gamma-rays from 
$\isotope{44}{Ti}$, 
$\kappa_{\gamma}(\isotope{44}{Ti}) =
0.040$, compared to $\kappa_{\gamma}(\isotope{56}{Co})= 0.033$. In our
calculations we 
use the the same opacity for \isotope{44}{Ti} and \isotope {56}{Co}.

In the case of a central energy source the total deposited energy may be
calculated from,
\begin{eqnarray}
4\pi \sigma_{\gamma} J_{\gamma} (r) & = & [(1 - e^{-\Delta
\tau_{\gamma,56}(r)}) e^{-\tau_{\gamma,56}(r)} L_{\gamma,56}+
\cr
 & & \cr
 & & (1 - e^{-\Delta \tau_{\gamma,57}(r)}) e^{-\tau_{\gamma,57}(r)}
L_{\gamma,57} + \cr
 & & \cr
 & & (1 - e^{-\Delta \tau_{\gamma,44}(r)}) e^{-\tau_{\gamma,44}(r)}
L_{\gamma,44}] /n_{\rm ion}(r), \cr
 & & \cr
 & & 
\label{eqn:depc}
\end{eqnarray}
where the index 56 refers to \isotope{56}{Co}, 57 to
\isotope{57}{Co}, and 44 to \isotope{44}{Ti}, the quantity $\tau
(r)$ is given below in equation (\ref{eq:tau}), $\Delta \tau(r) = 
\sigma (r) n(r) \Delta r$ is the optical depth of the shell of
thickness $\Delta r$, and $n_{\rm ion}(r)$ is the number of particles
(i.e. ions) at $r$.

If the radioactive isotopes are distributed throughout the
core region, the calculation of $J_{\gamma}$ is more
complicated.
Knowing the
luminosities for the radioactive isotopes (eqs.~[\ref{eq:lum56}] -
[\ref{eq:lum44}]), and their 
distribution (from the explosion model) we find the emissivity as a
function of radius, $ j_{\gamma}(r)$
We then calculate the total deposited energy from
\begin{eqnarray}
4\pi \sigma_{\gamma} J_{\gamma} (r) & = & 4 \pi \frac{1}{2} \int_{-1}^{1}
\int_{0}^{l_{\rm max}(\mu)} [\sigma_{\gamma,56} (r) 
j_{\gamma,56}(l) e^{-\tau_{\gamma,56}(l)} \cr
 & & \cr
 & &  + \sigma_{\gamma,57} (r) j_{\gamma,57}(l) 
e^{-\tau_{\gamma,57}(l)} \cr
 & & \cr
 & & + \sigma_{\gamma,44} (r) 
j_{\gamma,44}(l)  e^{-\tau_{\gamma,44}(l)}] ~  dl~ d\mu. \label{eq:depdis}
\label{eq:deps}
\end{eqnarray}
For each $r$, inside the core ($r < R_{\rm core}$),  we integrate over all
angles, $\mu = \cos\theta$, integration interval [-1,1]. For each
$\mu$ we integrate along a path, $l$, from $r$
($l=0$) to the border of the core region, $R_{\rm core}$ ($l=l_{\rm max}$). 
Outside of the core ($r > R_{\rm core}$) the integration interval for
$\mu$ is [-1,-$\sqrt{r^2 - R_{\rm core}}/r$]. The optical depth, $\tau$, in
equation (\ref{eq:depdis}) is given by
\begin{equation}
\tau (l) = \int_0^l \sigma (l') n(l') dl', \label{eq:tau}
\end{equation}
for each  isotope.

In general we can write
\begin{equation}
J_{\gamma} = \frac{L_{\gamma} D_{\gamma}}{16 \pi^2 R_{\rm core}^2},
\end{equation}
where $R_{\rm core}$ is the core radius.
Limiting forms of $D_{\gamma}$ are discussed in KF92.

\subsubsection{Line Transfer}
\label{sec-lintran}
In our calculations we take the radiative transfer of the lines into
account by using the
Sobolev approximation (Sobolev 1957, 1960, Castor 1970). The
supernova ejecta is expanding
homologously at the times we are interested in.
Therefore, each mass element, with mass coordinate $ M$, has
a constant velocity, $v(M) =$ constant, or $r(M) = v(M) t$,
where $t$ is the time since core
collapse. The expansion velocity is much larger than the thermal
velocity of the matter, and the Sobolev approximation is therefore
valid for an individual, well separated line.  
With this approximation the line transfer becomes a purely
local phenomena, either the 
photon is re-absorbed locally or escapes the medium.
A photon emitted in a certain transition may, however,  be absorbed by a
line with a longer wavelength at a different point in the 
ejecta. This line blocking is
important in the UV, and we hope to treat this in a subsequent
paper (see also Fransson 1994, Li \& McCray 1996, Houck \& Fransson
1996).
The effect of this has been checked by
comparing calculations for SN1993J (Houck \& Fransson 1996) with and
without these processes included, where it was found that the
strongest lines from dominant ions with high ionization
potentials, like O I, were only slightly affected, while elements with
low ionization potentials and low abundances, like Na I, are
sensitive to the UV-field. The electron density and temperature were
only marginally affected.
We test the
sensitivity of our results to the UV-field by having the option of 
switching it on or off.
 This is discussed in Paper II. 
Our crude treatment of the
radiative transfer is one of the major weaknesses of our calculations.

The Sobolev escape
probability, $\beta_{\rm s}$, can be written
\begin{equation}
\beta_{\rm s} = \frac{1 - e^{-\tau_{\rm s}}}{\tau_{\rm s}}.
\end{equation} 
Here  $\tau_{\rm s}$ is the Sobolev optical depth for a particular
transition, $ij$ ($i<j$),
\begin{equation}
\tau_{\rm s} = \frac{A_{ji} n_i \lambda_0^3 }{8\pi} \frac{g_j}{g_i} t (1 -
\frac{g_i}{g_j} \frac{n_j}{n_i}) 
\label{eq:sobolev}
\end{equation}
In addition of being re-absorbed in the line itself, a line photon may
also be destroyed by continuum absorption, i.e., photoionization. 
Note that the Sobolev expression for the escape probability (eq.
[\ref{eq:sobolev}]) is independent of assumptions of the line profile (e.g.,
Rybicki 1984).

\subsubsection{Continuum Destruction}
\label{sec-contdest}
The continuum destruction probability, $\beta_{\rm c}$, has in the Sobolev
approximation been calculated by
Hummer \& Rybicki (1985) for the Doppler case and complete redistribution,
\begin{equation}
\beta_{\rm c} \approx  {k_{\rm C} \over k_{\rm L}} F(\xi,{k_{\rm C}
\over k_{\rm L}}).
\end{equation}
In this equation $ k_{\rm C}/k_{\rm L}$, is the ratio between the continuum
opacity, $k_{\rm C}$, and the line core opacity, $k_{\rm L}$, and $\xi
= 1/\tau_{\rm s}$.
For a Doppler profile Hummer \& Rybicki (1985)  approximate
$F(\xi,{k_{\rm C} \over k_{\rm L}})$ by
\begin{equation}
 F(\xi,{k_{\rm C} \over k_{\rm L}}) \approx 2 \{-\ln \surd\pi({k_{\rm
C} \over k_{\rm L}} + 
2\xi[-\ln\surd\pi({k_{\rm C} \over k_{\rm L}} + 2\xi x_0)]^{1/2})\}^{1/2},
\label{eq:phdest}
\end{equation}
where $x_0 \approx 2$ is the frequency displacement from the line center
in thermal Doppler widths, $(\nu - \nu)/\Delta \nu_{\rm D}$.
 The assumptions made for this estimate  are that
$ k_{\rm C}/k_{\rm L} \ll 1$ and $\xi \ll 1$. 

For strong resonance lines the damping wings may become important. In this case the continuum
destruction probability above is modified. 
The destruction probability in the
Voigt case with $\it complete$ redistribution has for a static geometry been
calculated by Hummer (1968), giving expressions and tabulations for the function
$F$ above. For
$k_{\rm C}/k_{\rm L} \lesssim 0.1 a$ and $a \lesssim 0.1$, where $a$ is the damping parameter, Hummer gives the approximation 
\begin{equation}
 F(\xi,{k_{\rm C} \over k_{\rm L}}) \approx \left({\pi~a ~k_{\rm
 L}\over k_{\rm C}}\right)^{1/2}.
\label{eq:phdest2}
\end{equation}
For $k_{\rm C}/k_{\rm L} \gtrsim 0.1 a$ equation (\ref{eq:phdest}) gives an adequate
approximation.

The validity of complete redistribution versus partial redistribution has been
discussed extensively for solar conditions in the case of $\La$, Ca II
H \& K and Mg II (e.g.,  Mihalas 1978, and references therein). It is
there concluded that
at densities lower than those prevailing in the outer chromosphere,
partial redistribution is a better approximation to the line profile 
in the damping wings of the lines. Because the densities are in our case 
considerably lower than this, it is likely that partial
redistribution 
is a more realistic assumption for the strong resonance lines from the
ejecta. 
 The continuum
destruction has for partial redistribution been discussed by Chugai (1987), who finds the
approximation 
\begin{equation}
 F(\xi,{k_{\rm C} \over k_{\rm L}}) \approx 1.87~\left({a ~k_{\rm
 L}\over k_{\rm C}}\right)^{1/4}.
\label{eq:phdest3}
\end{equation}
This expression takes  only scattering in the wings into account,
and does not apply to the Doppler core, where complete redistribution,
as given by equation (\ref{eq:phdest}),
is  adequate.

The most important lines for which partial redistribution applies  are
$\La$, He I 
$\wl 584$, Ca II H and K, O I $\wl 1302$, and Mg II $\wl 2800$, as
well as the Fe II resonance lines. As
will be seen in Paper II, the choice of destruction
probability has important effects for especially the He I $\wl 584$
line and the branching to the $\wl 2.058 ~\mu$m
line. 
The effects of this are discussed in Paper II.

In our calculations we use the effective transition probability,
$A_{ji} \beta_{ji}$ where the escape probability $\beta_{ji}$ includes
both the Sobolev escape probability, and the continuum
destruction probability,
\begin{equation}
\beta_{ji} = (1-\beta_{\rm c}) \beta_{\rm s} + \beta_{\rm c}.
\end{equation}

\subsubsection{Photoionization and Photoexcitation/deexcitation}
\label{sec-phie}
In our calculation of the ionization balance we include
photoionization of all the ions with ionization potential smaller
than 30 eV. Where data is available, we also include photoionization
from excited levels in our multi-level atoms.
Lines, recombination emission, two-photon emission
and free-free emission may contribute to the photoionization.

We assume that only the continuum emission (recombination, two-photon,
and free-free emission) contribute to the
photoexcitation/deexcitation. 
Line photons destroyed by continuum absorption (see previous section)
are added to the photoionization of the different elements, according
to their fractional opacity.
References for the photoionization
cross sections are given in the appendix.

\subsubsection{Charge Transfer and Penning Ionization}
\label{sec-CT}
Charge transfer reactions may be important in determining the ionization
stages for the elements (Fransson 1994, Swartz 1994). This is especially the case for  trace
elements.   
The ionization balance of the dominating element in a zone is, however, 
not affected much by charge transfer reactions. Therefore, the
total electron fraction is rather insensitive to these reactions.
References to the sources of data are given in the appendix.

In modeling the helium emission from SN 1987A the Penning ionization process,
\begin{equation}
\begin{array}{lllllllll}
\HeI~(2^3S) & + & \HI & \rightarrow & \HeI~(1^1S) & + & \HII & + & e^-
\end{array}
\end{equation}
may be important for de-populating the excited level $2^3S$ of \HeI~
(Chugai 1991, Li \& McCray 1995). We include this process in our model.

\subsection{The Explosion Model}
\label{sec-explosionmodel}
As input to our calculations we require a model for the composition and mass
(or equivalently, the density) as a function of   velocity. The assumption
of homologous expansion relates the radius of a mass element to the
velocity, $r(M) = v(M) t$. Explosion models for SN 1987A have been
calculated by Nomoto \& Hashimoto (1988), Woosley (1988), Arnett (1987), and  Hashimoto, Nomoto,
\& Shigeyama (1989). These models are all one-dimensional
calculations. However, several groups have found that the
core is unstable to hydrodynamical instabilities, causing mixing of the
various burning shells, as well as modifying the density distribution
(e.g.,  Fryxell, M\"{u}ller,
\& Arnett 1991, Hachisu \etal 1992, Herant
\& Benz 1991, 1992). This mixing should be
macroscopic, with little mixing on the microscopic scale (Fryxell
\etal 1991). As
emphasized by e.g.  Fransson \& Chevalier (1989) and Li, McCray \&
Syunyaev (1993), macroscopic mixing, rather than microscopic,  can
have a major influence on the observed spectrum. 

In our calculations we assume an overall
spherical symmetry, with shells of different compositions and masses.
Unfortunately, most models which treat the nucleosynthesis in sufficient
detail are completely one-dimensional, 
with shells containing the heavy elements close to the center, and with continuously decreasing
mean atomic weight as one proceeds outward in radius.
Conversely, multi-dimensional models start with artificial initial
perturbations, and do not treat the nucleosynthesis in detail.
Because of limited
computing power it is also not yet realistic to include multi-dimensional
effects in the calculation of the spectrum, especially with regard to
the radiative transfer. 

\begin{figure}
\plotone{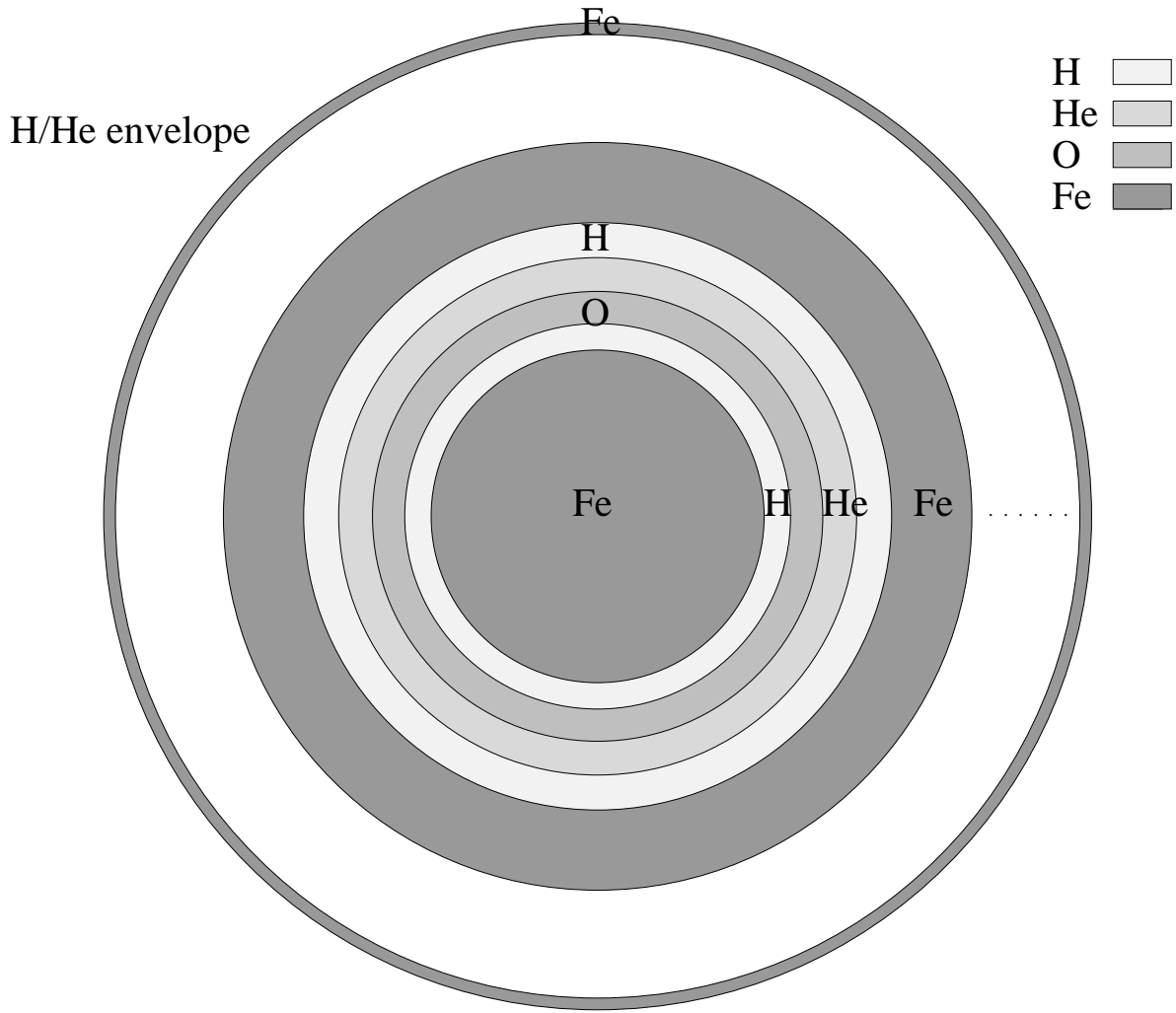}
\caption{Schematic figure showing the model structure of the ejecta,
where the thickness of the zones correspond to the filling factors of
the shells. We have for clarity omitted the O -- C and Si -- S zones,
as well as the various zones in the envelope. In the white zone
between the second and third iron zones we have an additional set of
O, He, H, Fe, O, He, and H zones. The inner hydrogen zone and the
outermost iron zone, which break the regularity, have been added to produce the correct line
profiles.  \label{fig:shells}}
\end{figure}

For these reasons we mimic the macroscopic mixing by using the results
of the one-dimensional explosion models with regard to the
nucleosynthesis, but not in density and velocity. Instead,
we divide the ejecta into a core region, containing the metal-rich
region, and where mixing is determining the distribution in velocity,
and an envelope, containing most of the hydrogen, as well as a major
part of the helium, and where mixing
is less important.
To  simulate
the structure of the multi-dimensional models in the core we divide
this into a number of
shells with varying compositions, placed in an alternating order.  
The density, $\rho_i$, of each of these shells is determined by the total mass of
the component in the core, $M_i$, and its  filling factor, $f_i$, so
that 
\begin{equation}
\rho_i = {3~M_i \over 4 \pi~ f_i~V_{\rm core}^3~t^3} . 
\end{equation}
Each component is then divided into a number of shells, usually three,
all having this density, 
placed uniformly within the core. In some cases (see below) we
determine the location, and masses of these shells, so that  a best
fit to the  line profiles is obtained.
A schematic representation of the structure, including only a few of
the zones, is given in Figure \ref{fig:shells}.

Based on line widths of [O I], Mg I], and [Fe II] lines we take the
velocity of the core to be $2000 \kms$. The structure of the envelope 
is based on the line profiles of $\Ha$ and the He I $\wl 2.058~ \mu$m
line (Paper II). 
In our calculations we use a maximum expansion velocity
for the envelope of $6000 \kms$. 

To represent the most important burning zones in the
post-explosion ejecta we include six different abundance
components.   
The dominant elements (by number)  are 
 Fe -- He, Si -- S, O -- Si -- S/O -- Ne -- Mg, O -- C, 
He, and H. Abundances are
either from the 10H model  (Woosley \& Weaver 1986, Woosley 1988) or
the 11E1 model (Nomoto, \& Hashimoto 1988, Shigeyama, Nomoto, \& Hashimoto 1988, Shigeyama \&
Nomoto  1990, Hashimoto \etal 1989). 
The abundances used, together with the masses of the different
components, for the 10H and 11E1 models are presented in Tables 
\ref{tab:10H} and \ref{tab:11E1}.
The sodium abundance is not given in the 10H model. We estimate
this by scaling it from the magnesium abundance, so that 
 the integrated $X(\isotope{23}{Na})/X(\isotope{24}{Mg})$ ratio is that
given by the 11E1 model.
This gives $X({\rm Na}) \approx 8.2 \EE{-3} ~ X({\rm Mg})$. 
The mass of synthesized stable nickel, M(${}^{58}$Ni)$+$  M(${}^{60}$Ni) $=
0.006 \Msun$, is from Nomoto \etal (1991).
The abundances in the hydrogen-rich
component in SN 1987A are based on modeling of the circumstellar ring
 by Lundqvist \& Fransson (1996), with C:N:O = 1:5:4, by
number. For sodium, magnesium, argon, and calcium we use 0.4 times the
cosmic abundance 
in the hydrogen-rich zones, and 
LMC abundances of Fe (Russell \& Dopita, 1992).
The stable nickel abundance is determined from the Ni/Fe ratio in LMC from
Russell \& Dopita (1992). For stable cobalt, the solar abundance ratio, Co/Fe,
from Anders \& Grevesse (1989) is used. 
The explosion models only include synthesized elements.
Therefore, if the abundance given by the explosive model is smaller 
than the LMC abundance, we estimate it from the mass fraction 
abundance in the hydrogen-rich component. 
In the Fe -- He and Si -- S regions  cobalt is  radioactive and 
its abundance therefore changes with time. 
 
The main physical differences in the 10H and 11E1 models are the
different treatments of 
convection in the pre-supernova, and different ${}^{12}{\rm C}(\alpha,
\gamma){}^{16}{\rm O}$ rates.
 Woosley \& Weaver (1986) employ the Ledoux 
criterion, in combination with semi-convection and over-shooting, while Nomoto \&
Hashimoto (1988) use the Schwarzschild criterion.

The He -- C regions in  the 10H and 11E1 models
are similar both in mass and in abundances. 
The oxygen-rich zones, however, differ considerably between the two models.
First, the mass of
the outer O -- C region is much larger in the 10H model, compared to the
11E1 model,
$\sim 0.60 \Msun$ and $\sim 0.10 \Msun$, respectively. Secondly,
in the 10H model the remainder of the oxygen zone has silicon and
sulphur as the 
most abundant elements, with abundances $\sim 0.065$ and $\sim
0.039$, respectively. The oxygen zone in the 11E1 model is, on the
other hand, dominated by neon and magnesium with abundances of $\sim 0.11$
and $\sim 0.071$. Only the inner part contains large fractions of silicon
and sulphur. The mass of the O -- Si -- S zone is in the 10H model $\sim
1.2 \Msun$, while in the 11E1 model 
the O -- Ne -- Mg contains $\sim 1.8 \Msun$. 
The composition of the Si -- S zone is
similar in the two models, but the mass is in the 10H model $\sim
0.30 \Msun$, while it is in the 11E1 model only $\sim 0.10 \Msun$. 
A case of special importance is the mixing of calcium into the
oxygen zone at the time of oxygen burning in the 10H model. In the
11E1 model this mixing is nearly absent. The calcium abundance is
therefore much larger in the 10H  model than in the 11E1 model. 
This has important consequences for the
evolution of the observed \CaII~ lines, as will be further discussed
in Paper II.

The velocity distribution and filling factors of the different components require a special discussion.
Herant \& Benz (1992) find that the central 'nickel bubble' extends to $\sim$
$1600 \kms$, that the oxygen shell is located between $600 - 2500
\kms$, and the helium
shell between $700 - 2600 \kms$. The hydrogen envelope penetrates in
these simulations down to $\sim 600 \kms$. The exact numbers are sensitive
to both the progenitor model and the nature of the perturbations put
into the simulations, and the results should therefore only be taken
as indicative. However, a large filling factor of the iron
component, coupled with similar low filling factors of both the
oxygen, helium and hydrogen components inside of 1700 -- 2000 $\kms$
seem to be a generic feature. In the intermediate region between 2000
-- 3000 $\kms$ both the hydrogen, helium and oxygen components have
comparable filling factors. Outside of $\sim$ 3000 $\kms$ the dynamics
is less affected, and the one-dimensional models may be fairly
realistic. The quantitative properties of the hydrogen envelope,
however, depend on the progenitor model. 
The line profiles can here set interesting
constraints, as will be discussed in Paper II.

In addition to the hydrodynamical simulations, we use,
as far as possible,
the observational information available. 
Li,
McCray, \& Sunyaev (1993) have  studied the evolution
of the Fe -- Co -- Ni clumps in SN 1987A. As long as the clumps are opaque
to the
gamma-rays they absorb most of the
radioactive energy, causing the blobs to expand. The
result is a 'nickel bubble', encompassing much of the volume of the
core, with a low density compared to the rest of the core. Li \etal
estimated the filling factor for these clumps to $\gtrsim 0.30$, with a
favored value of $\sim 0.5$. Their filling factors can not be directly
compared to ours as  
they assumed a maximum velocity of
$2500 \kms$ for their Fe -- Co -- Ni clumps, while we assume a
maximum value of $2000 \kms$.
In our
calculations we use $f_{\rm Fe} = 0.55$ for the Fe -- He component. 
This agrees with the conclusions of Basko (1994).

We use a  filling
factor of $f_{\rm Si-S} = 9 \EE{-3}$ for the Si -- S component 
 in the 10H model, and $f_{\rm Si-S} = 3 \EE{-3}$ in the 11E1 model. 
These filling factors give the same density in the Si -- S as in  
the O -- Si -- S/O -- Ne -- Mg  regions, respectively. 

Spyromilio \& Pinto (1991), as well as Li \& McCray (1992),
find from the observed [\OI]$\wll$6300, 6364 doublet ratio
a local oxygen density of $N_{\rm O}(t) \approx 6.2 \EE{10}~ t_2^{-3}
{\rm cm^{-3}}$.  Assuming an
oxygen mass  $M_{\rm O}$
$\sim$ 1.3 $\Msun$, consistent with the values found from explosion
models, they find a filling factor of the oxygen clumps of $f_{\rm O} \approx$
0.1. 
Assuming the same density 
in the oxygen-rich regions we find  
$f_{\rm O-Si-S} = 0.06$, $f_{\rm O-C} = 0.03$ for the 10H model, and
$f_{\rm O-Ne-Mg} = 0.09$, $f_{\rm O-C} = 5.0 \EE{-3}$ for the 11E1 model. 

For the helium-rich zone we use $f_{\rm He} = 0.15$ within the core, and   
the filling factor used for the hydrogen within the core is
$f_{\rm H} = 0.20$. 
These estimates are based on the simulations of Herant \& Benz (1992).

For the hydrogen envelope we have tested the density structure of the
Shigeyama \& Nomoto (1990) 14E1 envelope model, as well as
our own parameterized  density structure, based on the line
profiles (Paper II). 
The latter model is given by 
\begin{equation}
\rho = 9.1\EE{-16}~\left({t \over 500 {\rm~ days}}\right)^{-3}~\left({V \over
2000 \kms}\right)^{-2} ~\rm g~cm^{-3}
\label{eq:dens_env}
\end{equation}
at $V > V_{\rm core}$.
This density is flatter than both the 11E1 and 14E1
models.

Because the deposition, and therefore the different line luminosities,
depend directly on the individual gamma-ray optical depths, we give
these at 500 days in Table \ref{tab:gamma}. For other epochs these
scale as $t^{-2}$. Note that positrons add an additional amount of
energy to the Fe -- He region, which dominates the energy input to
this region at late epochs.

\section{RESULTS}
\label{sec-results}
\subsection{Temperature Evolution}
\label{sec-temp}
\begin{figure}
\plotone{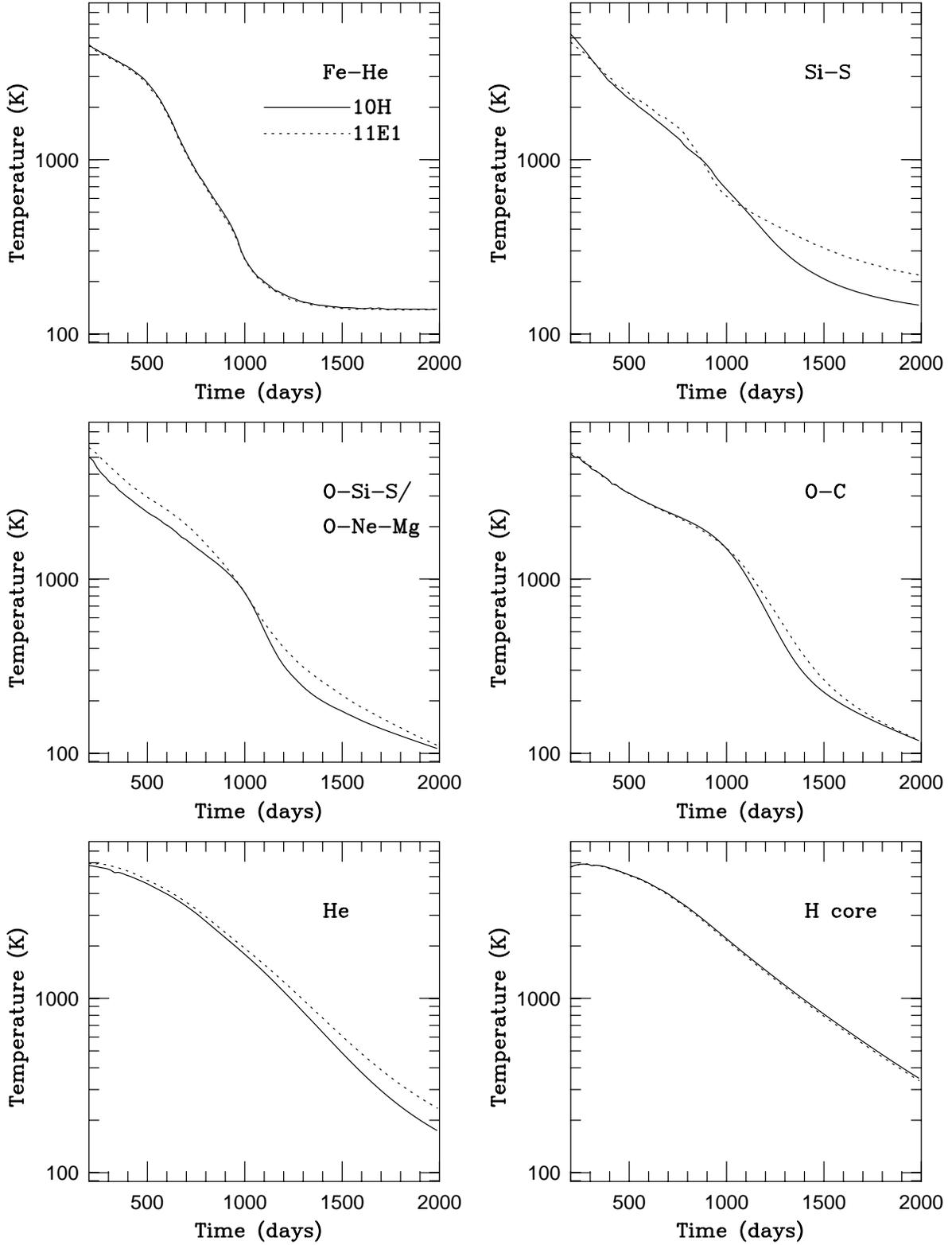}
\caption{The temperature evolution for a
representative shell for the different composition zones in
the core for the  10H (solid line) and
11E1 (dashed line) models.
  \label{fig:tempcore}}
\end{figure}

In Figure \ref{fig:tempcore} the temperature evolution for the different
components in the core are shown for the 10H and 11E1 model,
respectively. 

For the metal-rich shells, 
Figure \ref{fig:tempcore} shows an initial
slow decline, followed by a steep decrease of the temperature during a
few hundred days, after which the temperature levels off again. The
rapid decline in temperature is due to a thermal instability, and has 
been referred to as the IR-catastrophe (Axelrod 1980, Fransson
\& Chevalier 1989, Fransson 1994). 
The instability first occurs in the Fe -- He zone, followed by the Si --
S and the O --
Si -- S / O -- Ne -- Mg zones, and finally by the O -- C zone. 

This instability is not seen in the
helium, or hydrogen-rich regions. 
The reason for this is that at the time when  fine structure
cooling takes over in the oxygen-rich gas, at $\sim$ 1000 days, 
the cooling rate is a factor of $\sim 4$ lower in the helium-rich 
than in the oxygen-rich region. For hydrogen the corresponding factor is $\sim$ 75.
At the time when fine structure cooling in the helium and
hydrogen-rich 
regions becomes important, and
could cause an instability,
adiabatic cooling already dominates. The transition to the adiabatic
phase occurs at 
$\sim$ 700 -- 800 days in the hydrogen core, and at $\sim$ 900 -- 1000
days in the helium-rich core. When adiabatic cooling dominates  $T \propto
t^{-2}$. 

The temperature  in the time-dependent case is higher than in steady
state. The reason  is that energy from previous
stages, when heating, and therefore also radiative cooling were
important, is stored in the gas, 
and is then gradually lost in adiabatic expansion. In steady state a
decrease in the heating is directly balanced by cooling. The adiabat which
a given gas element follows is in the time-dependent case set by the epoch when radiative cooling
is equal to the adiabatic.

\begin{figure}
\plotone{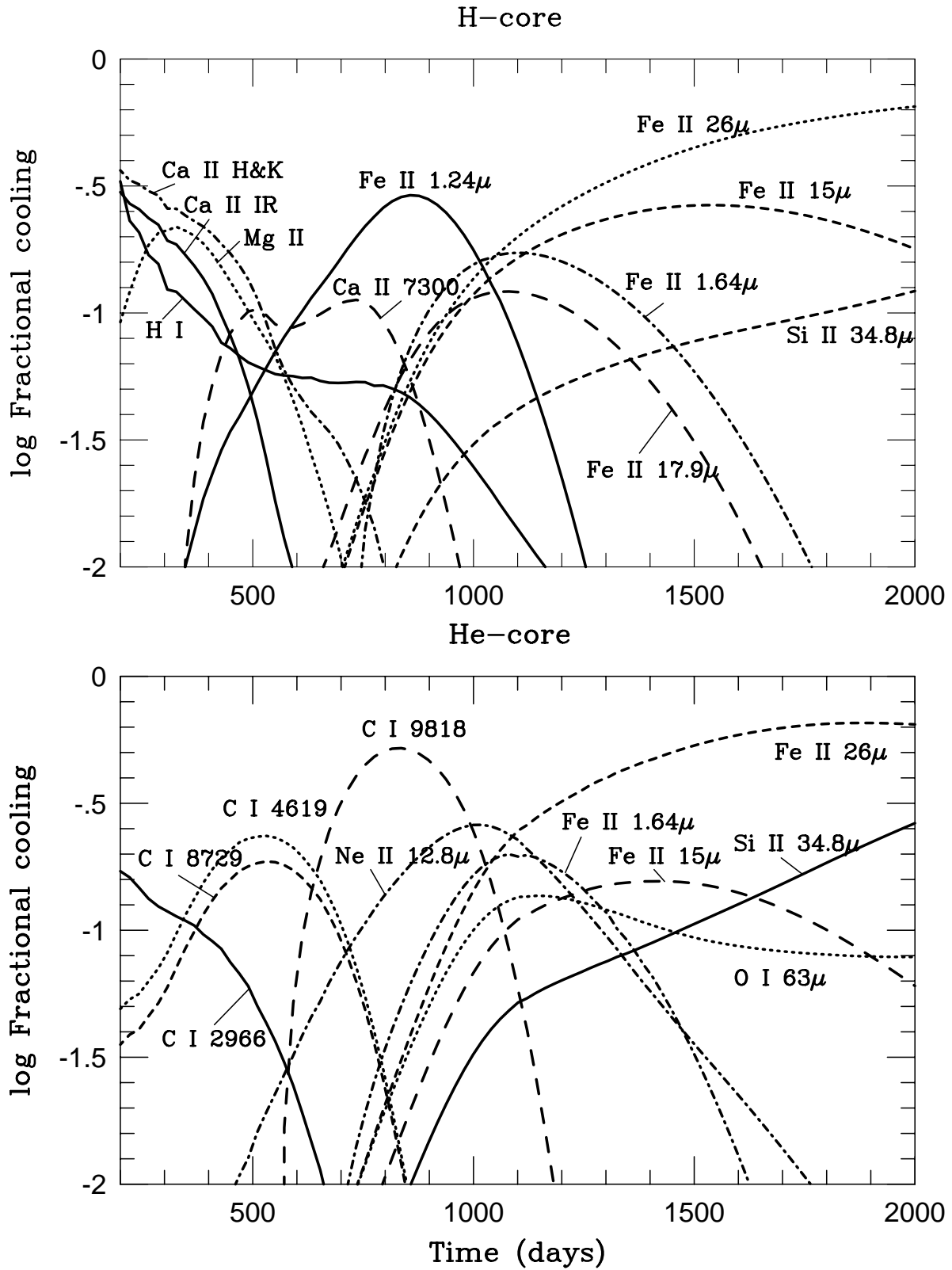}
\caption{Fractional cooling by the most important transitions in the
H-core zone and the He-core zone, as a function of time.
  \label{fig:cool_H_He}}
\end{figure}

\begin{figure} 
\plotone{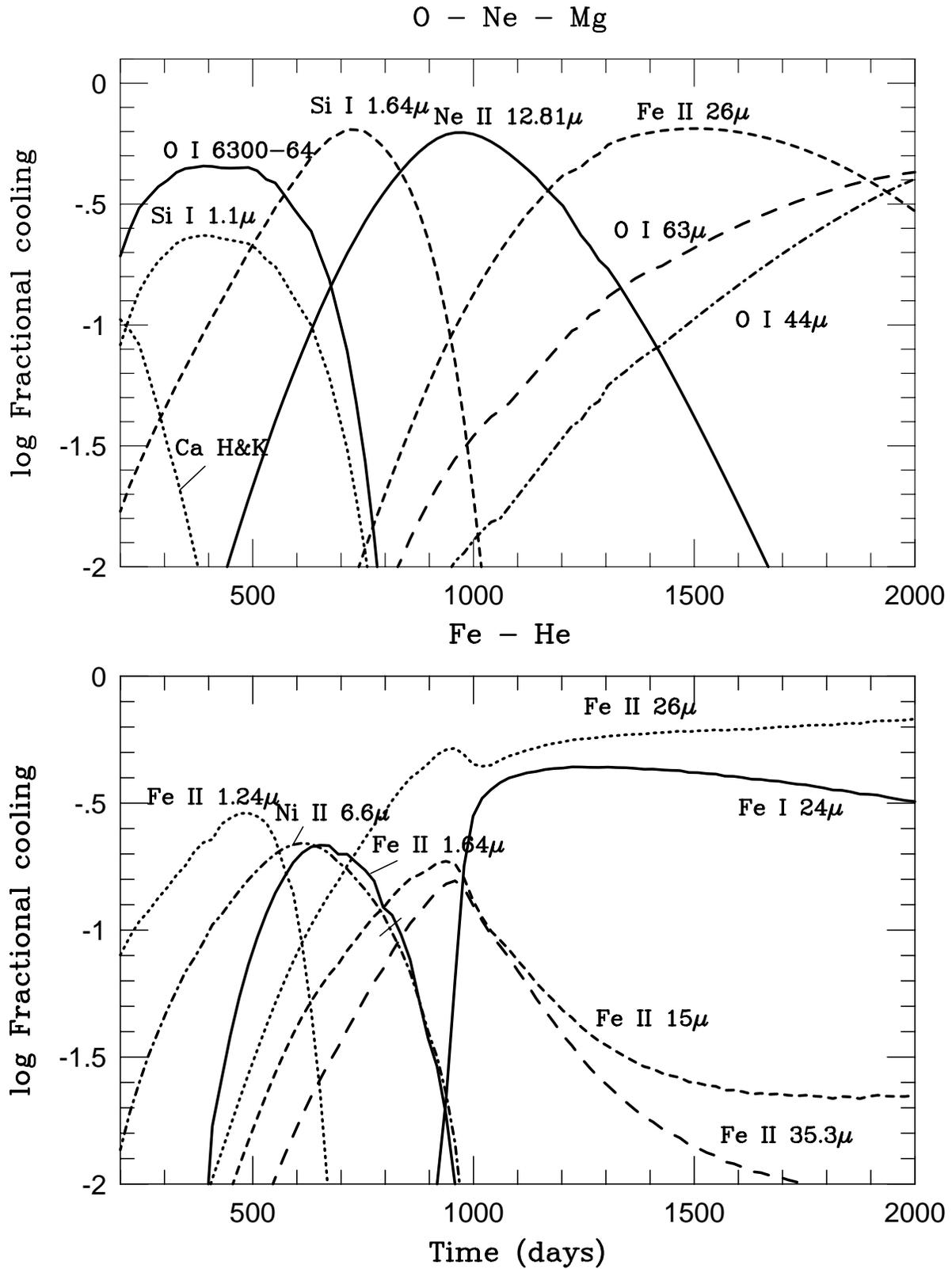}
\caption{Fractional cooling in the
O -- Ne -- Mg and the Fe -- He zones, as a function of time.
  \label{fig:cool_O_Fe}}
\end{figure}

\begin{figure}
\plotone{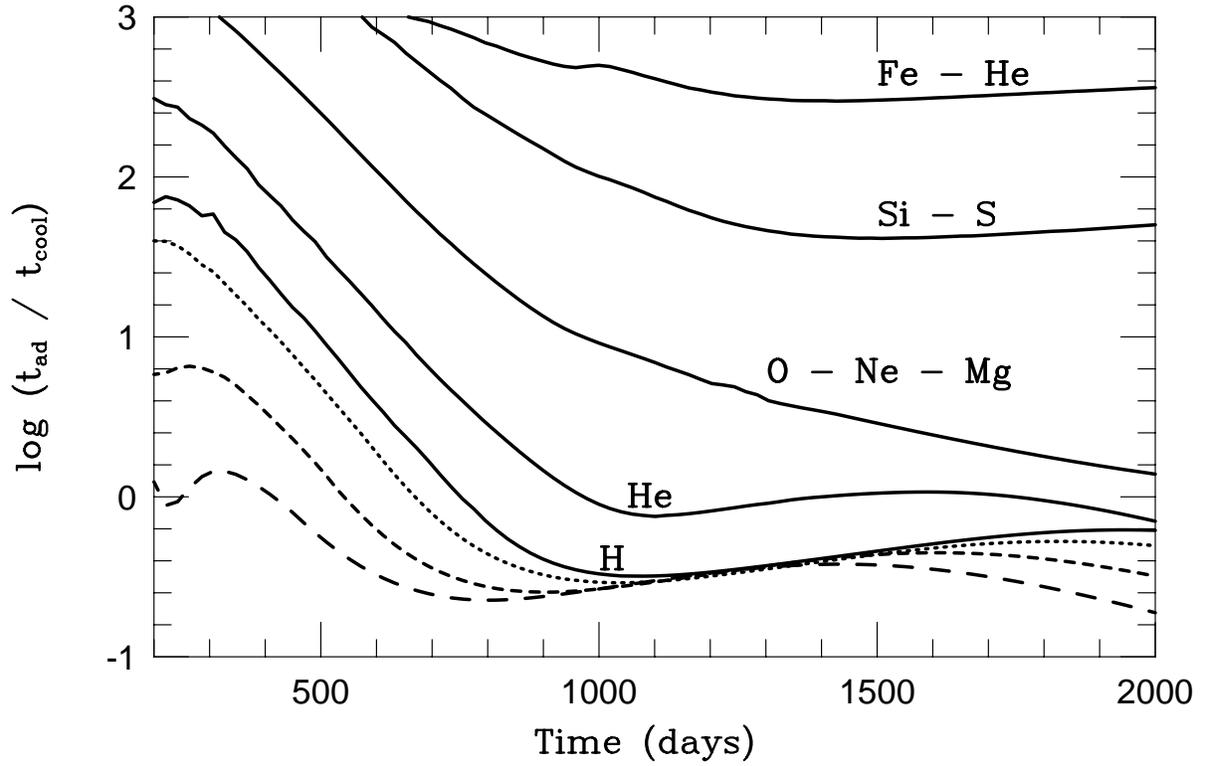}
\caption{Ratio of adiabatic to radiative cooling. The hydrogen curves
are from the top the hydrogen in the core (solid), and the hydrogen
envelope at 2000 $\kms$ (dotted), 2800 $\kms$ (short dashed) and 4100 $\kms$
(long dashed), respectively.
  \label{fig:adiab_cool}}
\end{figure}

Because the contribution from the different zones to a given line
reflects the temperature of this zone,
we now comment on the detailed evolution of each of these zones. 
To guide in this discussion we show in Figures \ref{fig:cool_H_He} and \ref{fig:cool_O_Fe}
the contributions to
the fractional cooling  from some of the most important transitions in
the H-core, the He-core, the O -- Si -- S, and the Fe -- He 
zones.  Note
that these figures give the collisional cooling of the gas (see
equation \ref{eq:lincool}), which is in
general not the same as the radiative losses. In particular,
transitions which are  
completely radiatively forbidden  may be strong
collisional coolants. An example is the $a{}^6D_{5/2}  -  a{}^6D_{9/2}
\wl 14.98 ~\mu$m transition in the ground multiplet of Fe II, which
contributes a comparable amount to the cooling as the observationally
important $a{}^6D_{7/2}  -  a{}^6D_{9/2}
\wl 25.99 ~\mu$m transition. 
The excitations in the $\wl 14.98 ~\mu$m transition mainly decay by
emitting line emission at $\wl 25.99 \mu$m and $\wl 35.35 \mu$m.
We also show in Figure \ref{fig:adiab_cool}
the relative importance of adiabatic cooling. 

\begin{figure}
\plotone{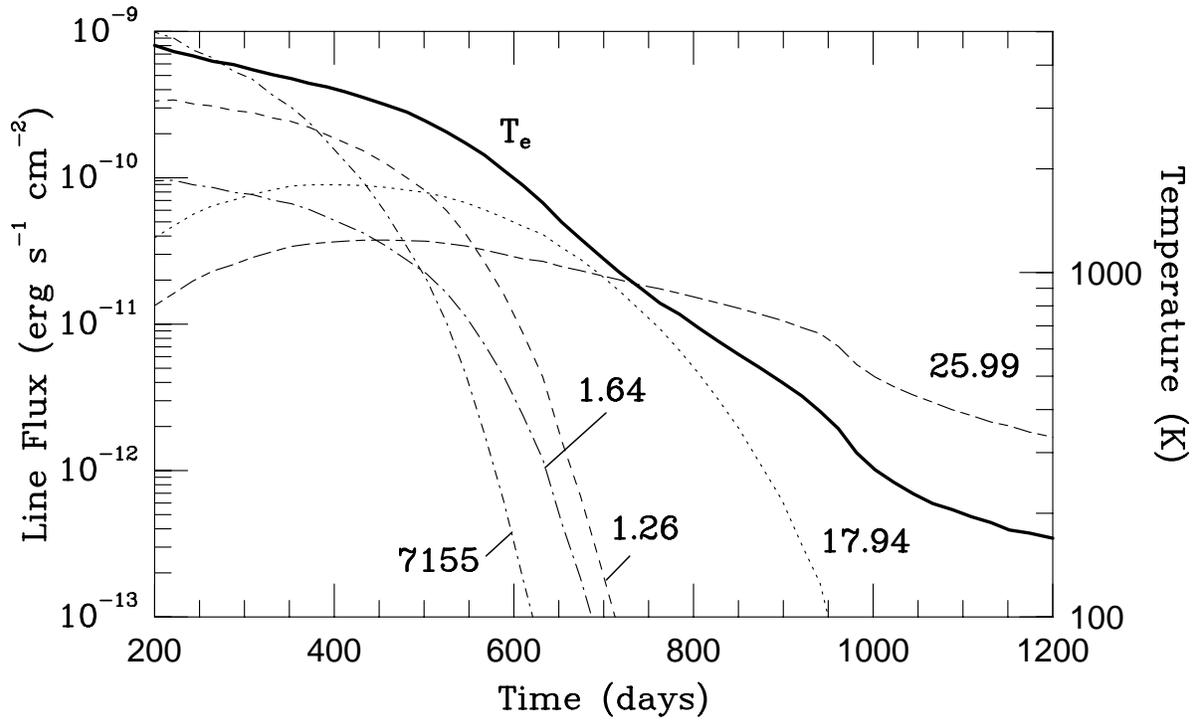}
\caption{The strongest Fe II lines from the iron core,
together with the temperature in the same region. Note the gradual
transition from optical to near-IR, to far-IR lines, as the
IR-catastrophe sets in at $\sim 600$ days.
  \label{fig:feii_fehe}}
\end{figure}

In the Fe -- He rich shells heating is dominated by non-thermal
heating of the free electrons at all times. A large fraction of this is due to the
deposition of positrons. At 200 days this is $\sim$ 45 \% of the non-thermal
heating, while at 1500 days the corresponding
number is $\sim$ 97 \%.
At most 20 \% (but
most of the time less than 10 \%) of the
heating is due to photoionization.
The cooling is at all times dominated by line cooling from  Fe
II.
The IR-catastrophe sets in at $\sim
 500$ days, when the temperature is $2700$ K, and is directly
reflected in the relative fluxes from the iron core (Fig.
\ref{fig:feii_fehe}).  
As the temperature falls, there is a gradual progression of cooling
from UV, to optical lines like $\wl 7155$, to near-IR, e.g., $\wll$
1.26, 1.64 $\mu$m, to far-IR fine structure lines.
At $t \gtrsim 500$ days,
Fe II $\wll$ 17.94, 25.99 $\mu$m
are the most important, and finally, at $t \gtrsim 750$ days, only the
$\wl$ 25.99 $\mu$m 
line contributes to the cooling. Once the IR-catastrophe has occurred,
the temperature is nearly constant at $\sim 140$ K. Because of the
efficient fine-structure cooling, adiabatic cooling is not important
before 2000 days.

From the observations of the [Fe II] $\wl$ 25.99 $\mu$m line we can get an
observational estimate of the temperature in the iron region, assuming
that the $\wl$ 25.99 $\mu$m line is the dominant coolant. This only applies
to the iron-rich gas and not to e.g. the hydrogen- or helium-rich
components. 
In the iron-rich gas the optical depth of the $\wl 25.99 \mu$m line is 
\begin{eqnarray}
\tau_{26\mu} & = & 28.5~f_{\rm Fe}^{-1}~\left({M_{\rm Fe} \over 0.07~\Msun}\right)~
\left({V \over 2000~\kms}\right)^{-3}~ \cr
 & & \cr
 & & \left({t
\over 500 ~{\rm days}}\right)^{-2}~(1 - e^{-554{\rm~K}/T}).
\label{eq:taufe26fe}
\end{eqnarray}
With a filling factor $f_{\rm Fe} \approx 0.5$ (Li, McCray \& Sunyaev 1993), and a
temperature $\gtrsim 200$ K (Fig. \ref{fig:tempcore}), the $\wl$ 25.99 $\mu$m line is
likely to be thick to at least $\sim 2000$ days. The critical density of the
$\wl$ 25.99 $\mu$m line is $\sim 2\EE4 {\rm~cm^{-3}}$, so the a$ {}^6$D$_{9/2}$
and  a$ {}^6$D$_{7/2}$ levels should be in LTE, and therefore radiate at
the blackbody rate,
\begin{eqnarray}
{dE \over dt dV} & =& {8 \pi hc \over \lambda^4~t~(e^{hc/\lambda k
T}-1)} \cr
&=& 2.55\EE{-12} ~ (e^{554{\rm~K}/T}-1)^{-1}~ \cr
 & & \left({t
\over 500 ~{\rm days}}\right)^{-1} {\ergs ~{\rm cm^{-3}}}.
\label{eq:dedvfe26}
\end{eqnarray}
(e.g., Fransson 1994). The heating rate in the iron-rich gas is
\begin{eqnarray}
{dE \over dt dV} &=& {\epsilon_{\rm H}~D_{\gamma} ~\rho \over 4
\pi R^2_{\rm core}} (\kappa_{\gamma,56} ~
L_{\gamma,56}+ \kappa_{\gamma,57} ~
L_{\gamma,57})\cr
&\approx& 0.75\EE{-9}~ f_{\rm Fe}^{-1}~
D_{\gamma}~\left({M_{\rm Fe} \over 0.07~\Msun}\right)^2~ \\
&&\left({V_{\rm core} \over 2000~\kms}\right)^{-5}~ 
\left({t \over 500 ~{\rm days}}\right)^{-5}~ \cr
&&e^{-t/111.3{}^d}~(1 + 1.2\EE{-3}~e^{t/155.6{}^d}) \nonumber
\label{eq:heat_fe} ~{\ergs ~{\rm cm^{-3}}}. 
\end{eqnarray}
We have taken a fraction $\epsilon_{\rm H} \approx 0.5$ going into heating (KF92),
and assumed a ${}^{57}$Co mass of $3.3\EE{-3} \Msun$.  At $t \gtrsim
900$ days the 
$\wl 25.99~ \mu$m line dominates the cooling (Fig. \ref{fig:feii_fehe}), and we can estimate the  
temperature in the gas by balancing the two expressions above.  Setting
$D_{\gamma} \approx 2.5$  (KF92) we obtain
\begin{eqnarray}
T_e  & \approx & 554 ~ 
\{ \ln [1 + 
1.4\EE{-3} \left({M_{\rm Fe} \over 0.07~\Msun}\right)^{-2}~
f_{\rm Fe}~ \cr
&& \left({V_{\rm core} \over 2000~\kms}\right)^{5}~ 
 \left({t \over 500 ~{\rm days}}\right)^{4} ~\cr
&&{e^{t/111.3{}^d} \over (1 + 1.2\EE{-3}~e^{t/155.6{}^d})}
] \}^{-1}      ~{\rm K}.
\label{eq:temp_fe}
\end{eqnarray}
At 900 days we find for $f_{\rm Fe} = 0.5$ that $T_e \approx 191$ K, at
1000 days $T_e \approx 140$ 
K, at 1200 days $T_e \approx 97$ K, and at 1500 days $T_e \approx 73$
K. These temperatures are about a factor of two less than those in
Figure \ref{fig:tempcore}. In spite of
this, equation (\ref{eq:temp_fe}) 
provides a good illustration of the sensitivity of the temperature to the
parameters involved. It also reproduces the leveling off of the
temperature in the Fe -- He core at $t \gtrsim 1000$ days (Fig. \ref{fig:tempcore}).  

The total luminosity in the $\wl 25.99 ~\mu$m line from the Fe -- He core is
\begin{eqnarray}
L_{26 \mu}& =& {32 \pi^2 ~h~c~f_{\rm Fe} V_{\rm core}^3~t^2 \over 3
\lambda^4~(e^{hc/\lambda k T}-1)} \cr
&=& 6.9\EE{36} f_{\rm Fe}~ \left({V_{\rm core} \over
2000~\kms}\right)^{3}~ \\
&& ~\left({t \over 500 ~{\rm days}}\right)^{2} ~
(e^{554{\rm~K}/T} -1)^{-1}  \ergs. \nonumber
\label{eq:lum26mu}
\end{eqnarray}
The observed luminosity at 636 days was  $\sim 6.0\EE{36} \ergs$ (Colgan \etal
1994). Multiplying this by a factor 1.7 to account for dust absorption,
and assuming $f_{\rm Fe} \approx 0.5$ gives a temperature of $1270$ K at
this epoch.  However, in Paper II we will show that there is likely to be a considerable
contribution 
to the flux of the $\wl$ 25.99 $\mu$m line from the hydrogen component. This temperature should
therefore be considered as an upper
limit. In our numerical model we find $T_e \approx 1520$ K at 636
days. Given the uncertainties in the filling factor and the dust
absorption this is probably consistent with this temperature. 

The temperature in the iron 
zone has also been calculated by Li, McCray \& Sunyaev (1993). 
In their calculations 
they consider clumps consisting of only iron, cobalt, and nickel. 
In nucleosynthesis
models  there is a significant fraction of helium 
($\sim 50 \%$ by number), microscopically
mixed with the iron, as a result of the alpha-rich freeze-out. 
In our calculations we find that, even for X(He) $\approx 0.5 - 0.7$,
helium has a negligible effect on the conditions in the iron core. The
reason  is that only $\sim$ 10 \%  of the energy is deposited in
helium, and $\sim$ 90 \% in the iron peak elements.
The temperature we find agrees well with that found by Li \& McCray.

In the Si -- S shells the temperature evolution differs somewhat
between the two models. The instability sets in earlier in
the 11E1 model, at $\sim 800$ days, compared to $\sim 900$ days for the
10H model. The temperature at $t \gtrsim 1000$ days is higher by a
factor of up to $\sim 1.4$ for the 11E1 model, because of a
higher positron deposition in this model. 
Non-thermal heating dominates at all epochs, but photoionization
of mainly \CaI, \SiI, and \SI~ contribute up to as much as 40 \% of
the heating.
The positron contribution  to the non-thermal heating is 10 -- 20 \%
at 200 days, and more than 70 -- 90 \% at 2000 days. 
Important coolants in these shells are \SiI, \SI, \CaII, Fe I -- II,
and Ni I -- II.

Also in the O -- Si -- S shells  in the 10H model 
and the O -- Ne -- Mg shells in the 11E1 model the temperature
evolution differs somewhat.
The instability sets in somewhat earlier in the 11E1 model, at $\sim$ 800
days compared to $\sim$ 1000 days for the 10H model.
We find that the temperature is slightly lower in the O -- Si -- S
zones in 10H, compared to that in the O -- Ne -- Mg zones of 11E1.
At 2000 days the temperature is $\sim$ 100 K in both models.

Calculations of the temperature in the oxygen core, with and without CO, have been made by
Liu \& Dalgarno (1995). In their CO free case they find
a temperature of 
$\sim$ 2200 K at 800 days, which is considerably higher than ours.
The reason is likely to be found in the differences between our models in  
the composition and in the trace elements included. 
Li \& McCray (1992) estimate the oxygen  temperature  from
observations of the [O I] $\wll$ 6300, 6364 emission doublet to T
$\approx$ 4100 K at 200 days, and T $\approx$ 2900 K at 500 days,
in fair agreement with what we find.

The temperature within the O -- C region decreases
from $\sim$ 4000 -- 5500 K at 200 days, to $\sim$ 2700 K at 600 days, 
 to $\sim$ 1500 K at 1000 days,
and to $\sim$ 110 K at 2000 days. However, the O -- C zone is  favorable
to CO formation. 
Because we neglect the very efficient cooling by CO, our
temperatures are probably over-estimated by a large factor.
Liu \& Dalgarno (1995) find that for the CO emitting regions the temperature
is constant at $\sim$ 1800 K for the first year, after which it
decreases to $\sim$ 700 K at 800 days. We will discuss the
implications of this for the line emission in Paper II. 

In both of the oxygen-rich zones the temperature decreases
nearly adiabatically once the IR-catastrophe has occurred, in contrast to
the Fe -- He region (Fig. \ref{fig:adiab_cool}). The IR-catastrophe occurs
slightly earlier in the O -- Si -- S / O -- Ne -- Mg regions compared
to the O -- C region. However, inclusion of CO-cooling will affect 
this result.

For the helium shells within the core, line cooling is of the same
order as adiabatic cooling after $\sim$ 1000 days, resulting in a faster
decline in the temperature than the adiabatic $T \propto t^{-2}$ cooling. 
Outside of the core, adiabatic
cooling becomes important at a somewhat earlier time at $\sim$ 800 days.

\begin{figure}
\plotone{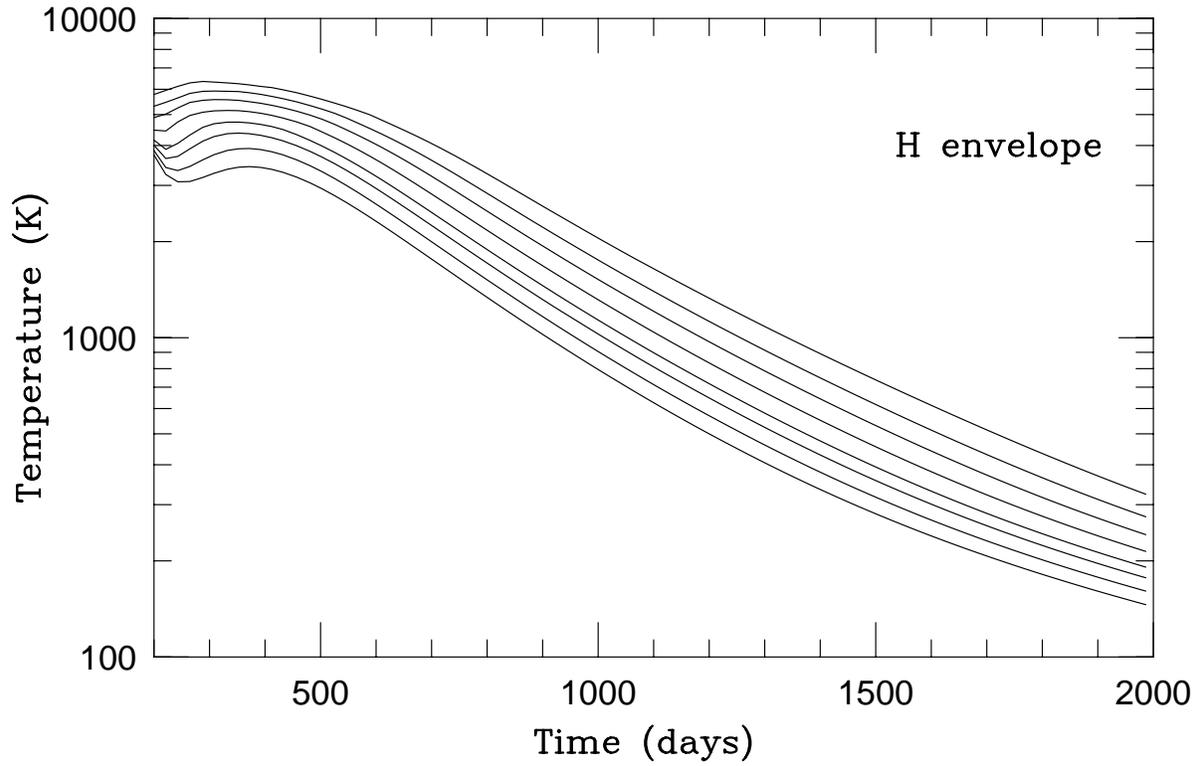}
\caption{The temperature evolution for the different shells in the
hydrogen envelope. The shell furthest in has the highest temperature.
The velocities of the shells are $\sim$ 2200, 2600, 3000, 3700, 4300, 
4700, 5250, 5750 $\kms$.
  \label{fig:tempenv}}
\end{figure}

In the hydrogen shells within the core,  
adiabatic cooling dominates line cooling after $\sim$ 800 days (Fig.
\ref{fig:adiab_cool}).
Figure \ref{fig:tempenv} shows the
temperatures for the different shells in the hydrogen envelope.
The
envelope density is in this, and in all other models, given by equation
(\ref{eq:dens_env}), unless otherwise stated.
The temperature in the envelope is at first constant, or even
decreasing,  up to $\sim$ 250 days. This decrease is sensitive to the
exact gamma-ray transfer, and therefore model dependent.  After
$\sim$ 250 days the temperature  decreases nearly 
adiabatically, with the highest temperature closest to the
core. Adiabaticity sets in first in the high velocity regions with the
lowest density.

\subsection{Ionization}
\begin{figure}
\plotone{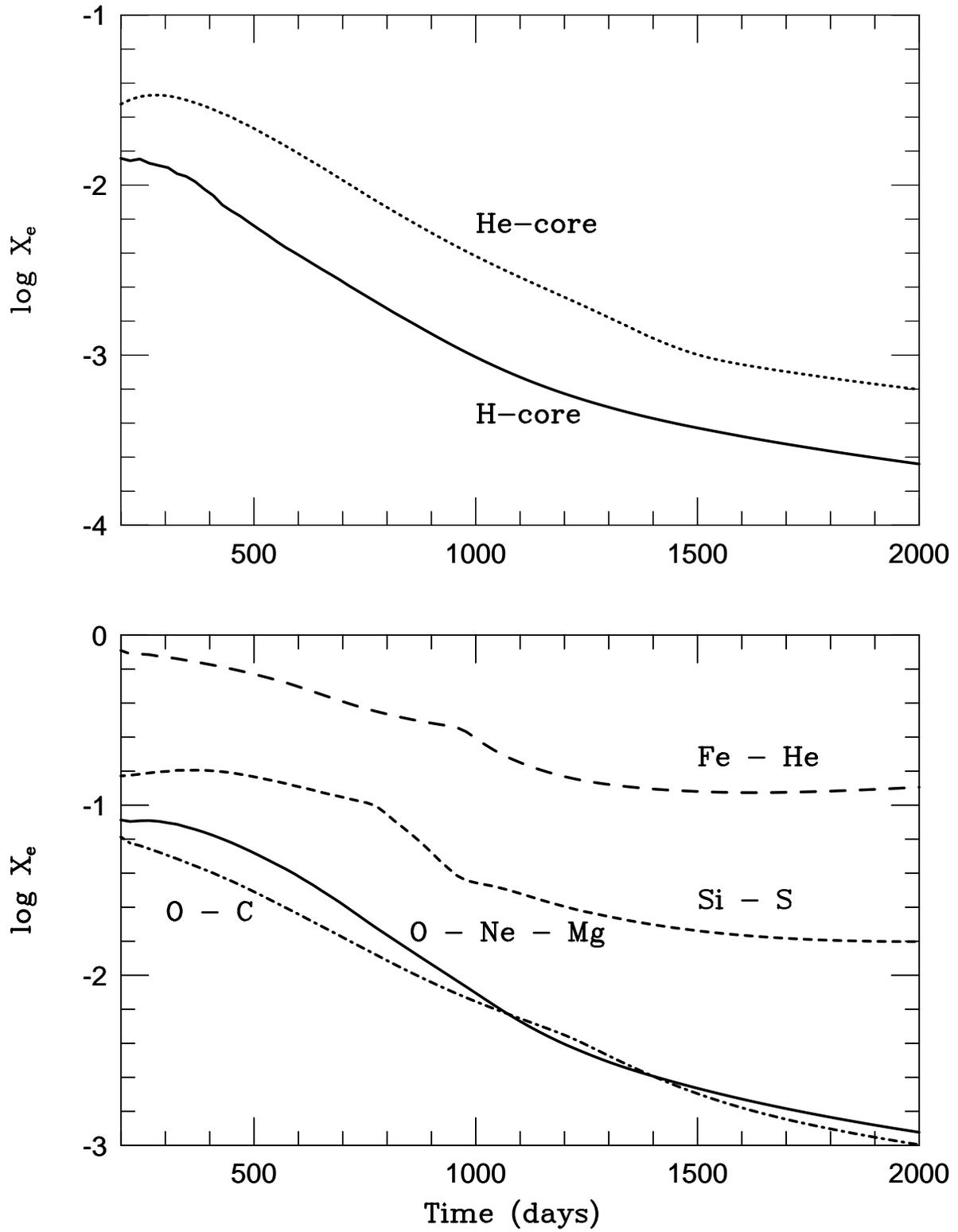}
\caption{The evolution of the electron fraction for 
representative shells for the different composition zones in
the core for the 11E1  model.
  \label{fig:xecore}}
\end{figure}

In Figure \ref{fig:xecore} we show the evolution of the ionization for the
same shells in the 11E1 model for which we previously showed the temperature
and cooling. 
The most interesting results are the gradual increase in the
ionization as we go to more and more advanced burning stages.  This is
mainly a result of the decrease in the number of ions for constant
density as we go to the heavier elements. Later than $\sim 1000$ days
the ionization in the iron core becomes nearly constant with time, in
contrast to the oxygen-rich zones.

\begin{figure}
\plotone{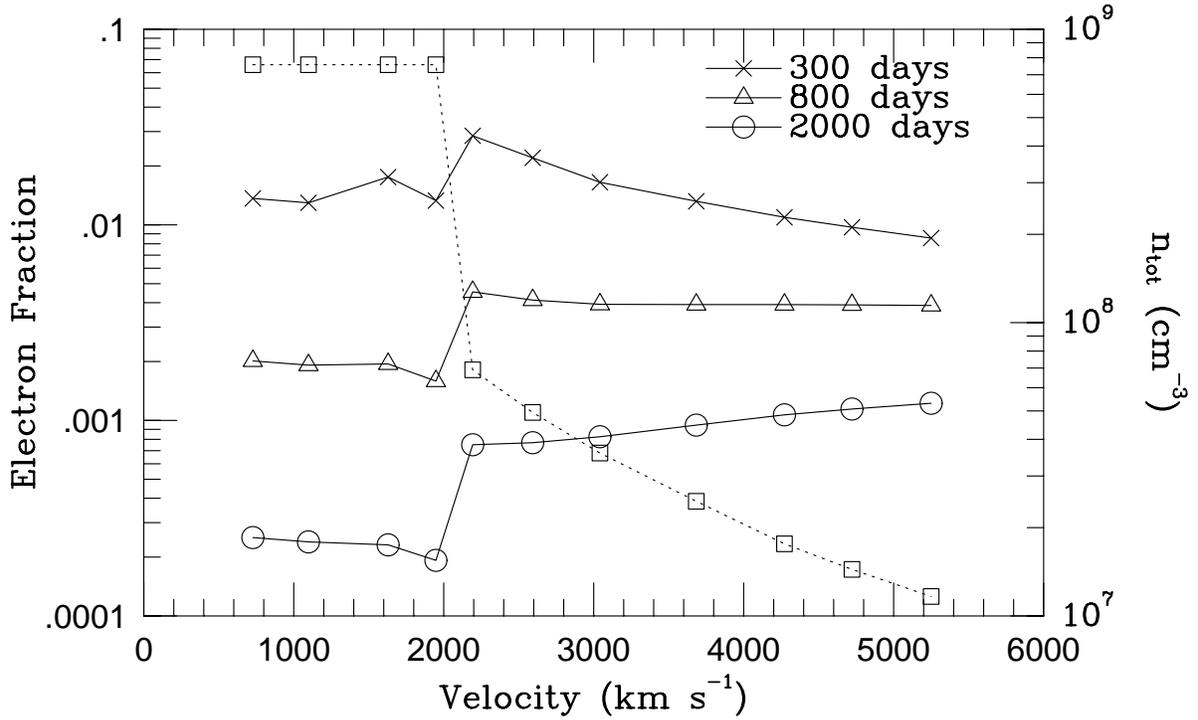}
\caption{Electron fraction as a function of velocity for the
hydrogen-rich regions at three different epochs. At 300 days
steady state is a good assumption, at 800 days the freeze-out effects
start to become important. Freeze-out effects are more pronounced
in the outer envelope region (V $\ge 2000 \kms$) where the density is lower.
Also shown, as the dotted curve, is the number density for the 
hydrogen-rich regions at 800 days.
  \label{fig:ionh}}
\end{figure}

Freeze-out of the ionization was discussed in connection to the
bolometric light curve in Fransson \& Kozma (1993). Because of the decreasing density the
recombination time scale may become longer than the radioactive decay
time scale.
The evolution of the ionization differs between the different composition
regions. The freeze-out effects 
sets in for all
compositions at $\sim 800 - 900 $ days. However, the degree of
freeze-out differs between the composition zones, and to a larger extent, between
the core and the envelope regions. 
The ratio of emitted energy, $L_{\rm em}$, to
deposited energy, $L_{\rm dep}$, is a useful measure of the degree of
freeze-out in the different regions (Fransson \& Kozma 1993). 
Within the metal-rich regions in the core 
$L_{\rm em}/L_{\rm dep} \sim 1.5$ is
reached. The corresponding number for the hydrogen-rich regions within
the core is $\sim 1.6 - 2.0$.
In the envelope, freeze-out sets in somewhat earlier (around
750 days for the regions furthest out) and a maximum of $L_{\rm em}/L_{\rm dep}
\sim 2.5$ is reached. To see the effect of the freeze-out on the ionization we show in Figure
\ref{fig:ionh} the electron fraction, $x_e$, as a function of velocity
for the hydrogen-rich shells at three epochs, 300, 800 and 2000 days. 
The lower ionization at V $\lesssim$ 2000 $\kms$ is caused by the higher
density assumed in the core region,
compared to the envelope (dotted line in Fig. \ref{fig:ionh}).
Because of the lower density, freeze-out
effects are increasingly pronounced in the outer regions of the
envelope. This is illustrated by the fact that the
decrease in ionization is smaller in these regions than in the inner
region of the envelope and within the core.

At 1000 days, the electron fraction and temperature in the envelope are
both a factor 
of $2 - 4$ higher in the time-dependent case, compared to steady
state. In the core, the increase in temperature and 
electron fraction is  $\lesssim 1.4$ for the
hydrogen and helium regions, while the difference in the metal-rich regions is
small. 

\section{EFFECTS OF UV-FIELD AND CHARGE TRANSFER}
\label{sec-ct2}
The UV-radiation field determines the ionization balance 
especially of ions with low ionization potentials and with small
abundances, like Na I, Mg
I, Si I and Fe I. Our model probably gives a  good
representation of the emissivity in the different lines and
continua. As has been discussed by several authors (e.g., Xu \& McCray
1991, Fransson 1994, Li \& McCray 1996), this radiation is 
severely affected by scattering by the many metal resonance lines
present. In addition, in the hydrogen-rich regions scattering by H$_2$
may be important (Culhane \& McCray 1995). The inclusion of these
effects is beyond the scope of this paper. In Paper II we comment on 
their importance. 

Besides the UV-field, charge transfer reactions are probably the most
uncertain point in our calculations. Reaction rates involving  hydrogen and
helium are  reasonably well known, because of their importance
in the interstellar medium. Reactions between metals are in
most cases uncertain, and in many cases completely unknown. In the
cases where rates are available these are usually motivated by their
importance in atmospheric physics. Unfortunately, extrapolations from
ions with similar atomic structure are usually impossible. In some
cases one can make an educated guess whether a certain reaction will be fast, if a
near resonance between the energy levels of the different ions is
present. This is e.g. the case for  O II + Ca I $\rightarrow$ O I + Ca II,
where the $5p {}^2P_{1/2}$ level in Ca II has an energy which together with the
ionization potential of Ca I is nearly equal to the O I ionization
potential. Consequently, the reaction rate is high,
$3.8\EE{-9}(T_e/10^3 {~\rm K})^{1/2} {~\rm
cm^{3}~ s^{-1}}$ (Rutherford  \& Vroom 1972). There are also cases where
this rule of 
thumb fails, and it is therefore of interest to estimate the influence
of these uncertainties for our calculations. 

An illustration of the sensitivity to charge transfer can be obtained by a simple
model with only two elements.
Of particular interest for the discussion of the O I recombination
lines in Paper II is charge transfer of Si I and O II in 
the O -- Si -- S component. 
 The Si I + O
II $\rightarrow$ Si II + O I
reaction can, like the O II + Ca I reaction, be expected to be fairly rapid,
based on the near resonance with the $3p {}^4P$ state in Si II, with
an energy difference of only 0.12 eV from the difference in ionization
potentials of O I and Si I. Each charge transfer reaction will then
give rise to a photon in the Si II] $\wll 2334 - 2350$ multiplet,
contributing to the UV-field.

 In steady state the
ionization balance is determined by (see
eq. [\ref{eq:ionbal}])
\begin{eqnarray}
\label{eq:ionbalct}
\Gamma_{{\rm OI}} n_{{\rm OI}} &=& 
n_{\rm SiI}  n_{\rm OII} \xi^{\rm CT}_{\rm OII+SiI} + \alpha_{\rm OII} n_e
n_{\rm OII} \cr
(\Gamma_{{\rm SiI}} + n_{\rm OII} \xi^{\rm CT}_{\rm OII+SiI} ) n_{{\rm SiI}} &=& 
 \alpha_{\rm SiII} n_e n_{\rm SiII}.
\end{eqnarray}
where $\Gamma_{{\rm OI}}$ is the non-thermal ionization rate of O I,
and similarly for Si I (eq. [\ref {eq:gammion}]). 
In addition, we have the number and charge conservation equations 
\begin{eqnarray}
n_{\rm O I} + n_{\rm O II} &=& n_{\rm O} \cr
n_{\rm Si I} + n_{\rm Si II} &=& n_{\rm Si} \cr
n_{\rm Si II} + n_{\rm O II} &=& n_e 
\end{eqnarray}

\begin{figure}
\plotone{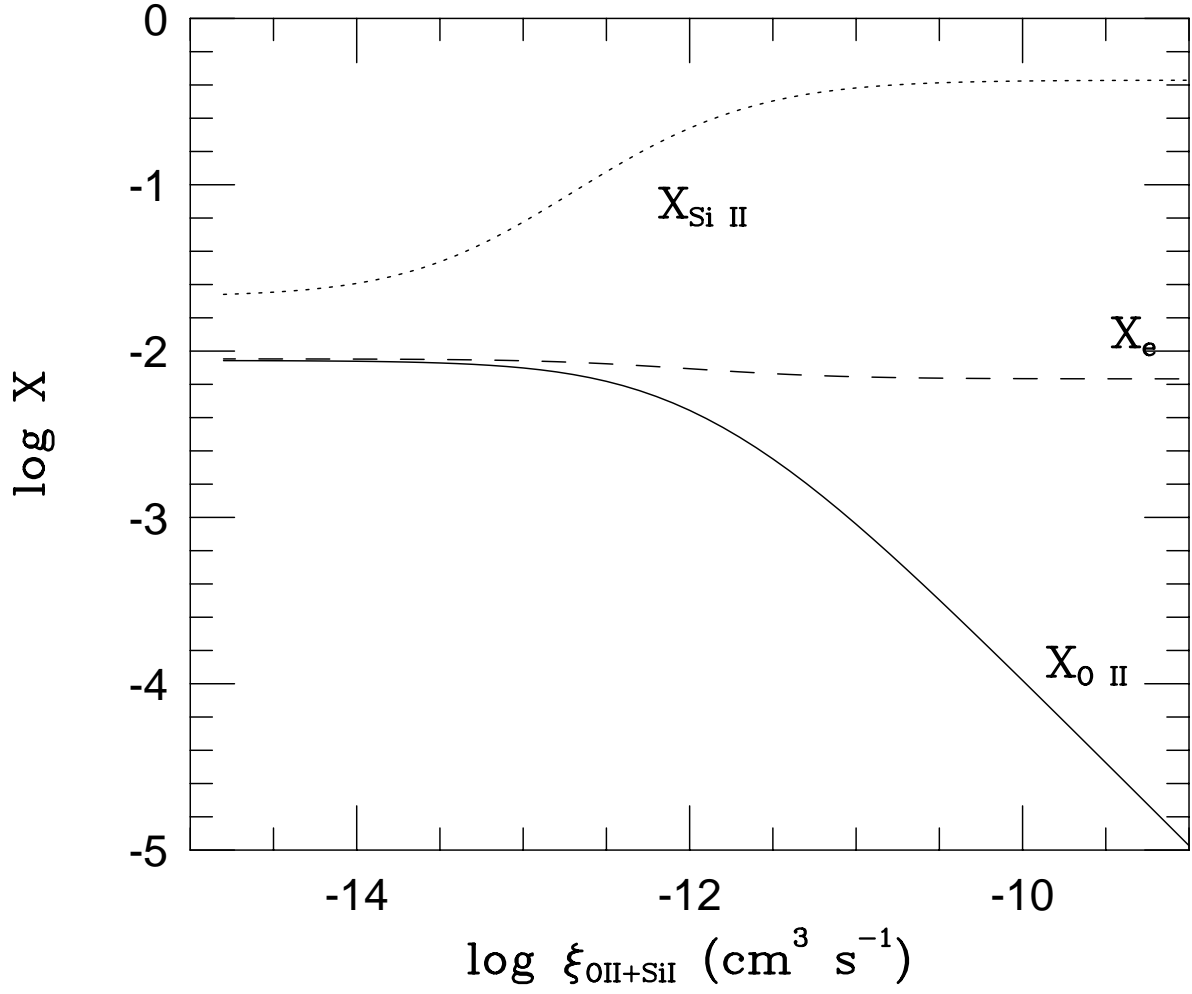}
\caption{Electron and ionic fractions of O II and Si II, as a function
of the Si I + O II $\rightarrow$ Si II + O I charge transfer rate. 
  \label{fig:ctosi}}
\end{figure}

These equations can easily be solved for a given epoch, and 
in Figure \ref{fig:ctosi} we show the
ionic abundances at 800 days for the case $X({\rm Si}) = 1.6\EE{-2}$ and
$X({\rm O}) = 0.98$ as a function of the charge transfer rate between Si I
+ O II. These 
abundances are similar to the 11E1 O -- Ne -- Mg region.
Although the
total electron fraction only varies by $\sim 25$ \% over the range $
\xi^{\rm CT}_{\rm OII+SiI} = 10^{-15}
- 10^{-8} {~\rm cm^{3} ~s^{-1}}$, the individual states of ionization
vary considerably. For $ \xi^{\rm CT}_{\rm OII+SiI} \gtrsim 10^{-11}{~\rm cm^{3}~
s^{-1}}$,  
basically all O II has disappeared, while the fraction of Si II
increases. This behavior is easy to understand, since when
$\xi^{\rm CT}_{\rm OII+SiI}~ n_{\rm Si I} / 
\alpha_{\rm O II}~n_e \gtrsim 1$ the O II fraction decreases proportional to
this ratio.  The fraction of Si II increases correspondingly. The
total electron fraction changes by a factor $\alpha_{\rm O II}/\alpha_{\rm
Si II}$. In this case $\alpha_{\rm O II}/\alpha_{\rm
Si II} \sim 1$, and only a minor change occurs. If $n_{\rm Si} \lesssim
n_{e}$ the neutral fraction of  silicon will decrease
substantially. If instead $n_{\rm Si} \gg n_{e}$, only a minor change
in the fraction of Si I will occur.
This illustrates a general characteristic of charge transfer. Only the minor
constituents, including ions of abundant elements, of a given zone are
affected be this process, in this case O II, Si I and Si II, but not O
I. This agrees with the result of Swartz (1994), where he finds 
that if the helium abundance is larger than the metal abundance the charge transfer
reactions do not greatly affect the ionization balance and level populations of helium.

We have tested this simple model by assuming an O II + Si I charge
transfer rate of $10^{-9} {~\rm cm^{3} s^{-1}}$ in our full code. This
leads to a large decrease of the O II fraction, as expected. There is,
however, little change in the Si I or Si II fractions. The reason for
this is that another charge transfer reaction, Mg I + Si II
$\rightarrow$ Mg II + Si I has a large rate. Therefore,  Si
II only acts as a mediator of the charge transfer from O II to Mg
II. This illustrates the gradual transfer of the ionization to
the low ionization potential elements, as well as the general
importance of charge transfer for the low abundance ions
(including ionized states of abundant elements, e.g. O II).

\section{DUST COOLING}
\label{sec-dustcool}
The presence of dust in SN 1987A can have important consequences for
the line emission. Besides the  effect of absorption, which will be
discussed in Paper II, dust cooling may also influence the thermal
evolution of the regions where  dust has formed. Although a
self-consistent discussion of this involves a solution of the equations
describing the nucleation, which is beyond the scope of this paper, we can
see the qualitative effects of the dust by a simple model, relying on
previous studies of the dust formation in SN 1987A by Kozasa,
Hasegawa \& Nomoto (1989, 1991).

Kozasa \etal  show that the most likely monomers to form
are corundum, ${\rm Al}_2~{\rm O}_3$, and enstatit, ${\rm Mg}~{\rm
Si}~{\rm O}_3$.  ${\rm Al}_2~{\rm O}_3$ 
forms first at a temperature of $\sim 1600$ K, while ${\rm Mg}~{\rm
Si}~{\rm O}_3$ forms 
at a somewhat lower temperature, $\sim 1500$ K. While Kozasa \etal
mainly considered microscopically mixed cases, we think this is
unlikely, as already 
argued. Therefore, there is little overlap of regions with a
substantial abundance of iron and oxygen, and the formation of
${\rm Fe}_3~{\rm O}_4$ is likely to be inefficient. 
Here we 
consider mainly the formation of ${\rm Al}_2~{\rm O}_3$ and ${\rm
Mg}~{\rm Si}~{\rm O}_3$ in 
the O --  Ne -- Mg zone, where both magnesium, silicon, and aluminum are
 abundant, together with oxygen. 

To model the formation of these grains we simply assume that they form
at the condensation
temperatures above. Kozasa \etal (1991) find a radius for the ${\rm
Al}_2~{\rm O}_3$ 
grains of $\sim 10$ \AA, while the ${\rm Mg}~{\rm Si}~{\rm O}_3$
grains have a radius of 
$\sim 70$ \AA. The cooling rate per dust grain is, assuming that
$T_{\rm dust} \ll T_e$, (Dwek \& Arendt 1992)
\begin{equation}
\Lambda_{\rm dust} = 1.26\EE{-27}~ [{n \over A_{\rm mean}^{1/2}} +
({m_p \over m_e})^{1/2}~n_e]~a^2~T_e^{3/2} \ergs,
\end{equation}
where $a$ is the grain radius in \AA. The first term represents 
cooling by atoms and the second that by the free electrons.

The total amount of dust formed is uncertain. Here we assume,
consistent with Kozasa \etal,  that all aluminum in the O -- Ne -- Mg
zone goes into ${\rm Al}_2~{\rm O}_3$ and all silicon into ${\rm
Mg}~{\rm Si}~{\rm O}_3$ grains. In 
the 11E1 model, $X({\rm Mg}) \approx 7.1\EE{-2}$, and $X({\rm Si}) \approx
1.6\EE{-2}$. We take the aluminum abundance from the 11E1 model to be $8.3\EE{-3}$ by mass. The total
mass of the O -- Ne -- Mg zone is $1.8 \Msun$, giving $M({\rm Al}) =
1.5\EE{-2} \Msun$, and the 
mass of ${\rm Al}_2~{\rm O}_3$ is then $2.8\EE{-2} \Msun$ and the mass
of ${\rm Mg}~{\rm Si}~{\rm O}_3 \sim 0.17 
\Msun$. 
With a size of
$1.7~$\AA~ and $2.06~$\AA~ per key species (Kozasa \etal 1989), the total number of grains in the
two cases are therefore $1.8\EE{51}~(M({\rm Al}) / 1.5\EE{-2}\Msun)~(a
/ 10 {\rm ~ \AA})^{-3}$ and
$5.2\EE{49}~(M({\rm Si}) / 5\EE{-2}\Msun)~(a / 70 {\rm ~
\AA})^{-3}$, respectively. The total, optically thin
cooling rate for ${\rm Al}_2~{\rm O}_3$ is then
\begin{eqnarray}
\Lambda_{\rm dust}^0 &=&  1.1\EE{-13} ~\left({M({\rm O}) \over
2\Msun}\right) ~\left({M({\rm Al}) \over
1.5\EE{-2}\Msun}\right)~ \cr
& & \left({V_{\rm core} \over 2000
\kms}\right)^{-6}
  ~\left({t \over 500{\rm~  days}}\right)^{-6}
 ~\left({f_{\rm O} \over 0.1}\right)^{-2} \cr
 & & ~\left({a \over 10 {\rm~ \AA}}\right)^{-1}~ (1 + 171~ x_e)~T_e^{3/2}
\ergs~cm^{-3}, \cr
 & &
\end{eqnarray}
and for ${\rm Mg}~{\rm Si}~{\rm O}_3$
\begin{eqnarray}
\Lambda_{\rm dust}^0 &=&  1.7\EE{-13} ~\left({M({\rm O}) \over
2\Msun}\right) ~\left({M({\rm Si}) \over
5\EE{-2}\Msun}\right)~ \cr
& & \left({V_{\rm core} \over 2000
\kms}\right)^{-6}
  ~\left({t \over 500{\rm~  days}}\right)^{-6}
~\left({f_{\rm O} \over 0.1}\right)^{-2} \cr
&&  ~\left({a \over 70 {\rm~ \AA}}\right)^{-1}~ (1 + 171~
x_e)~T_e^{3/2} \ergs~cm^{-3}. \cr
& &
\end{eqnarray}
As the temperature falls the cooling will eventually exceed the
black-body limit, at which time the temperature stabilizes.

Assuming that  dust formation occurs when the temperature has
decreased to the condensation temperatures for the two species, the
${\rm Al}_2~{\rm O}_3$ formation takes place at $\sim 800$ days. 
At the time of dust formation in our model, 
immediately before the temperature drop, the ratio
of the dust cooling and line cooling is $\sim 50$. Therefore, as soon
as the dust forms it immediately cools the gas to a very low
temperature, at which the optically thick dust cooling
balances the heat input. 
Because of the
dramatic cooling, the
${\rm Mg}~{\rm Si}~{\rm O}_3$ formation occurs at the same epoch. 

A dust formation epoch at $\sim$ 800 days is
later than the observations indicate. 
Possible reasons for this
discrepancy could either be extra cooling in the oxygen-rich gas due to
e.g. molecular cooling by SiO, or the simplified assumption of
instantaneous dust formation at the temperatures above.
Other possible, and perhaps the most likely, sites for dust formation
are in the O -- C zone and in the iron-rich core. 
In the O -- C  region CO cooling could decrease the
temperature enough for graphite to form. This scenario requires that
not all carbon is locked up into CO. This is supported by the 
calculations done by  Liu \& Dalgarno (1995). They find a CO mass of $(1 -
2)\EE{-3} \Msun$, while the total carbon mass in the O -- C zone is $(1 - 10)
\EE{-2} \Msun$.   
A second possible site is the iron core, where the temperature is
$\lesssim 1500$ K at $\sim 600$ days. Clumping may further decrease
the temperature and thus the dust condensation epoch.
The dust formation is likely to occur at different epochs at
different positions in the ejecta, explaining the observed gradual
formation of the dust.

In the optically thick limit we can calculate the dust temperature by
balancing the gamma-ray energy absorbed in the O --  Ne -- Mg zone by the
dust cooling to obtain 
\begin{eqnarray}
T_{\rm dust} &=&  1770 ~\left({V_{\rm core} \over 2000
\kms}\right)^{-1/2}
 ~\left({t \over 500{\rm~  days}}\right)^{-1}\cr
&&~\left({f_{\rm cov} \over 0.4}\right)^{-1/4}
 ~\left({\Delta \tau_{\gamma, \rm O}(500 {\rm~ days}) \over
0.23}\right)^{1/4}~ \cr
& &\cr
& & e^{-t/445{\rm~  days}} \rm~   K.
\end{eqnarray}
We have here scaled the gamma-ray optical depth of the O --  Ne -- Mg zone by
the value in our models at 500 days (Table \ref{tab:gamma}).
At 600 days we obtain $T_{\rm dust} = 380$ K, and at 800 days 183
K. These values are in reasonable agreement with those determined by
Wooden  \etal (1993) and Colgan \etal (1994). 

Although certainly too simplified,  this  discussion is indicative of
the importance of the additional cooling where dust is formed. The
cooling rate we estimate above is directly proportional to the 
amount of dust assumed to be formed, as long as the gas is optically thin. Because of
the nearly full conversion of the key elements into dust, this should
probably be 
considered as an upper limit. A serious
discussion of the dust cooling requires a more physical model of the
dust nucleation, along the lines of Kozasa \etal

A consequence of this scenario is a depletion of the elements in the
oxygen region involved in the dust condensation. 
This has little
effect on oxygen. Both aluminum and silicon are by assumption
completely depleted from the gas phase in this region, and magnesium
is decreased by $\sim 20 \%$ in the 11E1 model. Observational evidence
for Si depletion, based on the [Si I] lines, has been prestented by
Lucy \etal (1991). It is, however, not trivial to separate the effects
of dust depletion and decrease in temperature on the strength of the
Si emission.

\section{CONCLUSIONS}
\label{sec-conclusion}
The temperature and ionization of the supernova ejecta determine the
evolution
of the observed emission lines. In this
paper we have
made a considerable effort to calculate these as realistically as
possible. Previous
calculations have made various simplifications, which limit their
applicability.
Fransson \& Chevalier (1987, 1989) neglected time-dependent effects, as
well as
the hydrogen-rich component. Li,  McCray \& Sunyaev (1993) and Liu \& Dalgarno
(1995) only
considered the iron and oxygen-rich zones, respectively, and also
neglected time-dependent effects. 
The only work comparable to ours is that by de Kool, Li \&
McCray
(1997), which we became aware of only after this work was completed. In
as far as
we have been able to compare, we agree very well with their results in
terms of
both temperature and ionization. This comparison of completely independent
calculations adds  to the confidence of our results, within the
assumptions of the model.

Except as input to the calculations of the line emission, our most important
results concerns the IR-catastrophe at 600 -- 1000 days in the metal
zones. 
Other results of special interest concern
adiabatic cooling later than 500 -- 800 days in especially the hydrogen and helium regions, and
the
freeze-out of the ionization in the hydrogen envelope. Although some
of these
effects have been noted previously (Fransson \& Chevalier 1987, 1989;
Fransson \& Kozma 1993), they are in this paper synthesized into a coherent
picture. In Paper II we show how the temperature evolution is
reflected in the line emission. In addition, both molecule and dust
formation are sensitive to the temperature. 
In our models dust is formed too late in the O -- Ne -- Mg zones.

In spite of our effort to include physics as realistically as
possible, there
are still some major deficiencies, in particular our treatment of the
UV-field, the neglect  of
molecular and
dust cooling, and uncertainties in the charge transfer
rates. Molecular cooling
can have important consequences, as demonstrated by Liu \& Dalgarno
(1995), but
is probably limited to the O - C zone, and the interface between the
oxygen and
silicon regions. As we will discuss in Paper II, the thermal line
emission from
the other regions would otherwise also be completely quenched. As our estimate
of the
dust cooling shows, a similar effect would occur where the dust
forms. In this
case, however, dust formation takes place where there is a strong
drop in
the temperature, because of the IR-catastrophe. The emission from the
metal-rich
zones is then dominated by non-thermal  excitation, and the line
emission is
therefore insensitive to the temperature.
These areas clearly deserve much further study.

\newpage

\appendix
\begin{center}
\section{APPENDIX}
\end{center}
\subsection{Charge Transfer}
Charge transfer reactions are important in determining the
ionization balance for the trace elements.
Unfortunately, rates are rather uncertain. The reactions we
include in our calculations are given in Table \ref{tab:ct}
together with references for the rates. 
Most  charge transfer rates with hydrogen are from Kingdon
\& Ferland (1996), who give analytical fits to all the rates. 
This includes a compilation of both 
already available rates from different sources (see
references therein), and newly calculated rates.  

The data in Prasad \& Huntress (1980) and Arnaud \& Rothenflug (1985)
are also compilations from various sources.
The rates from Kimura, Lane, Dalgarno, \& Dixson (1993) (H II + He I
$\rightarrow$ H I + He II), and Kimura \etal 1993 (He II + C I
$\rightarrow$ He I + C II) are low. 
For the reaction H II + Ca I $\rightarrow$ H I + Ca II we did not 
find any published data.
As an estimate we use the same rate as for O II + Ca I $\rightarrow$ O I +
Ca II.
 
Swartz (1994) estimates charge transfer reaction rates between helium
and metals, using the Landau-Zener and modified Demkov approximations. He also estimates the rates for the charge transfer
between excited states in He I ($2s ^1S$, $2p ^1P$, $2s ^3S$, and $2p ^3P$)
and metals. The accuracy of these are probably not better than a
factor of two to five. 
We therefore include the rates from Swartz separately, to examine their importance.

The rate for the Penning process, $7.5
\EE{-10} (T/300 K)^{1/2} {\rm cm^3 s^{-1}}$, is from Bell (1970).

\subsection{Model Atoms}
\subsubsection{H I}
For H I we use a model atom consisting of 30 levels. For the 5 lowest
$n$-states  all $l$-states are included. The higher $n$-states, up
to $n = 20$, are treated as single levels. 

We use expressions given by Brocklehurst (1971) for 
the transition probabilities between different $nl$-states
($A_{nl,n'l'}$). Only  transitions with $\Delta l = \pm 1 $ are assumed
to have A-values greater than zero. The one exception is the
two-photon transition,  
$2s - 1s$, with a transition probability of $8.23 ~{\rm s^{-1}}$.
When calculating transition probabilities between an upper $n$-state and
a lower $nl$-states, as well as between two $n$-states, the following
averages are used,
\begin{equation}
A_{n,n'l'} = \sum_{l=0}^{n-1} \frac{g_l}{g_n} A_{nl,n'l'},
\end{equation}
and
\begin{equation}
A_{n,n'} = \sum_{l} \frac{g_l}{g_n} \sum_{l'} A_{nl,n'l'} = \sum_{l'} A_{n,n'l'}
\end{equation}
where $g_l = 2(2l+1)$ is the statistical weight of the $l$-state, and
$g_n = 2n^2$ is the statistical weight of the $n$-state.
In thermodynamic equilibrium, when the $l$-states
are populated according to their statistical weights ($x_{nl} = x_n
\frac{g_l}{g_n}$), we have
\begin{equation}
x_n A_{n,n'l'} = \sum_{l} x_{nl} A_{nl,n'l'}
\end{equation}
and
\begin{equation}
x_n A_{n,n'} = \sum_{l,l'} x_{nl} A_{nl,n'l'},  
\end{equation}
motivating the expressions above.

Recombination coefficients to the $nl$-states are calculated using
expressions in Brocklehurst (1971). 
The total recombination coefficient for hydrogen is given by Seaton
(1959). These rates agree with the total recombination
coefficients by 
Verner \& Ferland (1996). In order to get the correct total
recombination coefficient we increase the recombination coefficient
to the highest ($n=20$) $n$-state, so that the sum of the recombination
coefficients to all the levels equals the total recombination coefficient.
This treatment is not correct for the highest levels, but should be a
good approximation for the lower levels.

Collisions with both electrons and protons between $l$-states for a given $n$-level are included for
the 5 lowest $n$-levels (Brocklehurst 1971, Pengelly \& Seaton
1964).
The total collisional excitation rates between different $n$-levels
($C_{nn'}$) are from Johnson (1972). To divide the total rates
into rates between different $l$-states we use the Bethe-approximation
(Hummer \& Storey 1987). For all $\Delta n > 0$ and $\Delta l = \pm
1$ we use,

\begin{equation}
C_{nl,n'l'} = \frac{A_{nl,n'l'}}{A_{nn'}} C_{nn'}
\end{equation}
and
\begin{equation}
C_{n,n'l'} = \frac{A_{n,n'l'}}{A_{nn'}} C_{nn'}.
\end{equation}
This approximation gives,
\begin{equation} 
x_n C_{n,n'} = \sum_{l,l'} x_{nl} C_{nl,n'l'},
\end{equation}
if the different $l$-states are assumed to be populated according to
statistical weights.

Photoionization cross sections for individual $nl$-states, for  $n > 2$, 
are calculated using the expressions given by Brocklehurst (1971), 
and for the higher
$n$-states by Seaton (1959).
Photoionization cross sections for $n=1$ and $n=2$ are from
Bethe \& Salpeter (1957).

To solve the level populations for the hydrogen atom using the Sobolev 
approximation complications appear due to overlapping lines. 
Transitions between two $n$-states but different $l$-states have the
same energy, and the lines may interact. However, because of the
degeneracy this scattering  is  local,  and the escape
formalism may still be used. This case has been treated by Drake \&
Ulrich (1980), and we use their expressions.

The treatment of an $n$-state
as a single level is correct for high enough densities, in which case the
$l$-states become populated according to statistical weights.
When calculating the line emission 
we 
recalculate the level populations for hydrogen using a 210 level atom
containing all $l$-states up to $n=20$. This is done separately from
the calculation of temperature and ionization, where we assume
$l$-mixing for $n \ge 6$.

For the hydrogen two-photon continuum we use the two-photon
distribution given by Nussbaumer \& Schmutz (1984).

\subsubsection{He I} 
We use a model atom  for He I,
consisting of 16 levels. The first 12 levels are  $1s ~ 
^{1}S$, $2s ~ ^{3}S$, $2s ~ ^{1}S$, $2p ~ ^{3}P$, $2p ~ ^{1}P$, $3s ~
^{3}S$, 
$3p ~ ^{3}P$, $3d ~ ^{3}D$, $4s ~ ^{3}S$, $4p ~ ^{3}P$, $4d ~ ^{3}D$, 
$4f ~ ^{3}F$. 
The remaining 4 levels are fictitious
levels, into which all the triplet levels up to $n=100$ have been combined.
Collision strengths, and
transition probabilities are from Almog \& Netzer (1989).
Verner \& Ferland (1996) have calculated recombination
coefficients, both  total and to the different excited states, for the
entire temperature range, 3 -- $10^{9}$ K, which we use.

In our model we use the direct recombination coefficients calculated
by Verner \& Ferland (1996) to all levels except to $2s ~ ^{1}S$ and $
2p ~ ^{1}P$ and the $4p ~ ^{3}P$ and $4f ~ ^{3}F$ states. The reason
for this is 
that the $2s ~ ^{1}S$ and $
2p ~ ^{1}P$ states are the only excited singlet levels included in
the model atom by Almog \& Netzer (1989). Including only direct
recombinations to these would therefore under-estimate the total
recombination rate, including higher states.  
We use the fit given by Almog \&
Netzer, which is, however,  uncertain at low
temperatures. This fact introduces some uncertainty in the He I
two-photon continuum  at low temperatures.

For the $ 4p ~ ^{3}P$ and $ 4f ~ ^{3}F$ levels we use rates from
Almog \& Netzer data in the 
temperature interval 10000 to 20000 K, extrapolating this to lower
temperatures by  $\alpha \propto T^{-0.5}$. This is in agreement with
the temperature dependence Bates (1990) finds for low temperatures.
Recombination coefficients to the four highest, fictious levels
are from Almog \& Netzer in the temperature interval 10000 --
20000 K. The temperature dependence is corrected for lower
temperatures, using $\alpha \propto T^{-0.5}$ for $T <$ 1000 K.
The 
rates of these four highest recombination coefficients were normalized to
achieve the correct total recombination coefficient. 

For the ground
state photoionization cross sections are from Verner \etal (1996),
while the excited state cross sections are from Koester \etal (1985). 
For the helium two-photon continuum, between the 2s $^1$S and the
1s $^1$S states, we use the
distribution given by Drake, Victor, \& Dalgarno
(1969).

In the Spencer-Fano calculations we include non-thermal excitations to
the following excited levels: 
$2s~ ^{3}S$, $2s ~ ^{1}S$, $2p ~ ^{3}P$, $2p ~ ^{1}P$, $3s ~ ^{3}S$,
$3p ~ ^{3}P$, $3d ~ ^{3}D$. References for the cross sections are
given in KF92.

\subsubsection{O I}
Our \OI~ atom consists of 13 levels, $2p^{4} ~ ^{3}P_2$, $2p^{4} ~
^{3}P_1$, $2p^{4} ~ ^{3}P_0$, $2p^{4} ~ ^{1}D$, 
$2p^4 ~ ^1S$, $3s ~ ^5S_o$, $3s ~ ^3S_o$, $3p ~ ^5P_o$, $3p ~ ^3P_o$, 
$4s ~ ^5S_o$, $4s ~ ^3S_o$, $3d ~ ^5D_o$, $3d ~ ^3D_o$. Transition probabilites and collision
strengths  are from Bhatia \& Kastner (1995). 
We include non-thermal excitations from the ground state to 
all excited levels. References for the cross sections used are
given in KF92. 
For the fine structure transitions within the $^{3}P$ state 
A-values and 
collision strengths for O I are from Berrington (1987).

The total radiative recombination coefficient is from 
Chung, Lin, \& Lee (1991). 
Recombination coefficients to the individual levels are
from Julienne,  Davies \& Oran (1974), renormalized by a factor 1.38 to
give a total rate equal to that of Chung \etal 

Photoionization cross sections are from Verner \etal (1996) for the ground
state. For the excited states the data are from Dawies \& Lewis (1973).

\subsubsection{Ca II}
The six-level model atom for \CaII~ consists of the   $4s ~ ^{2}S$, $3d ~
^{2}D$, $4p ~ ^{2}P$, and $5s ~ ^{2}S$ levels.
A-values are from Ali \& Kim (1988), Zeippen (1990), and Wiese
\etal (1969).
Collision strengths are from Burgess \etal (1995), except
between the sublevels of 3d, which are from 
Shine (1973).
Ground state photoioinization cross sections are from Verner \etal
(1996). For the excited states data are from Shine (1973), while 
recombination rates are from Shull \& Van Steenberg (1982). 

\subsubsection{Fe I-IV}
Fe I-IV are treated as multilevel
atoms with 121 levels for \FeI, 191 levels for \FeII , 110 levels for
\FeIII, and 43 levels for \FeIV. A-values for \FeI~ are 
from Kurucz \& Peytremann (1975) and from Axelrod (1980), and collision strengths
from Axelrod (1980). The accuracy of the latter are probably low. 
For \FeII, A-values and collision strengths are from the Iron
Project (Nahar 1995, Zhang \& Pradhan 1995), and from Garstang
(1962). Additional A-values are from Kurucz (1981). 
For \FeIII~ and \FeIV~  data come from Garstang (1957) and
Kurucz \& Peytremann (1975), Berrington \etal (1991). 
Total recombination rates are from Shull \& Van Steenberg
(1982), and the 
fractions of the radiative recombination going to the ground state are
 from 
Woods, Shull, \& Sarazin (1981). 

For \FeI~ and \FeII~ the total photoionization cross sections are
from Verner \etal (1996). Only photoionizations from the first five
levels (i.e., the ground multiplet) are included for  \FeI~ and
\FeII. The same cross section is assumed for all levels in the
ground multiplet.

It is highly time consuming to calculate the level populations
for the large Fe I-IV ions. In order to speed up the calculations  
we reduce the number of levels in these, if possible. The first time steps
of the calculation always use 'full' iron-atoms. At later epochs, the
populations of the 'full' atoms are calculated every few days. The
highest levels, which in these calculations together contribute less
than 0.1 \% of the total 
emission or cooling/heating, are then excluded in the following time
steps. The error in the total emission introduced by this procedure is
therefore never larger than $\sim 0.1 \%$.

\subsubsection{Ni I-II and Co II} 
We include Ni I, Ni II, and Co II with a small number of
levels in order to model the [Ni I] $\wl$ 3.119 $\mu$m, 
[Ni I] $\wl$ 7.505 $\mu$m, 
[Ni II] $\wl$ 6.634 $\mu$m, [Ni II] $\wl$ 10.68 $\mu$m, and the [Co II]
$\wl$ 10.52 $\mu$m lines.
We assume that the ionization balances for cobolt and nickel are
coupled to the ionization balance of iron by charge transfer (Li,
McCray, \& Sunyaev 1993). 
Due to lack of good atomic data we do not
include non-thermal excitations or recombinations in calculating the
line emission. 
Therefore,  lines arising from these atoms
are due only to thermal, collisional excitations by electrons. 

Our Ni I  atom has seven levels, consisting of the three
lowest multiplets, $a ^3F$, $a ^3D$, and $a ^1D$. Transition
probabilities between levels within the $a ^3F$ and the $a ^3D$ multiplets
 are from Nussbaumer \& Storey (1988), and transition
probabilities between $a ^1D$ and $a ^3D$  from Garstang (1964).   

The model atom for Ni II has eight levels,  consisting of the three
lowest terms, $^2D$, $^4F$, and $^2F$. Transition probabilities
are from Nussbaumer \& Storey (1982), and
collision strengths are from Bautista \& Pradhan (1996).
For Co II we only include the lowest multiplet, $a ^3F$, consisting of
three levels, with transition probabilities from Nussbaumer \& Storey
(1988).

Unfortunately,  no collision strengths are available for Ni I and
Co II. Axelrod (1980) estimated collision strenths from the empirical 
relation 
\begin{equation}
\label{eq:omega}
\Omega_{i,j} = C g_i g_j
\end{equation}
where $g_i$, and $g_j$ are the statistical weights for the levels and $C$ is
a constant.
The constant was determined 
by a comparison with calculations by
Garstang, Robb \& Rountree (1978) for Fe III, and found to be $C = 4
\EE{-3}$ for transitions with $\lambda < 10 ~\mu$m. 
For infrared transitions with $\lambda > 10 ~\mu$m he adopted a higher
value of the constant, $C = 2\EE{-2}$. 
For the lowest 16 levels in Fe II 
we have checked this relation 
with the new Opacity Project collision strengths (Zhang
\& Pradhan 1995), and found $C = 0.046$.
In order to test equation (\ref{eq:omega}), we calculated collision
strengths for Ni II using both  $C = 4\EE{-3}$, and $C = 0.046$,
and compared to the collision strengths from Nussbaumer \&
Storey (1982).  The Ni II data are best reproduced by the constant
from Axelrod, while our larger $C$ gives a factor of 10 too large collision
strengths. 
Therefore the constant from Axelrod gives acceptable values for Ni II,
while it under-estimates the collision strengths for Fe II by a factor
of 10.

Li, McCray, \& Sunyaev (1993) proposed an empirical formula for the
collision strengths, based on known transitions in Fe II. In their
formula the collision strength depends on the transition probability,
and the energy difference between the levels, as well as on the
statistical weight of the upper level. We do not find any clear
correlation between transition probabilities, energy differences and
collision strengths.

Given the result of the tests above, we feel that it is not possible
to estimate the unknown collision 
strengths empirically to any degree of certainty. We therefore simply
put the unknown 
collision strengths equal to 0.1 and await new atomic data.

\subsubsection{Other Ions}
References for the collisional ionization cross sections used for
calculating the non-thermal ionization rates, are given in KF92.
For the photoionization cross sections  we use the analytical fits
given by Verner \etal (1996).

Verner \& Ferland (1996) give analytical fits to the radiative
recombination coefficients in the temperature interval 3 K to $10^{9}$
K for recombinations towards H-, He-, Li- and Na-like
ions. We use their data for recombination of He III, Mg III, and Na II. 
Radiative recombination coefficients from Gould (1978), and
low-temperature dielectronic recombination 
coefficients from Nussbaumer \& Storey (1983, 1986) are used for 
 recombination of C II, C III, N II,
O III, Mg II, Si II, Si III.
 We use rates from Shull \& Van Steenberg (1982)
for  the Ne II, Mg II, S III,  
Ar II, Ca II recombinations.
For S II we take  radiative recombination rates from Gould (1978). 

To calculate the cooling and line emission, 
the level populations for a number of model atoms are 
calculated, in steady state.
C I, N II, O III, and S III are treated as six-level atoms, and
C II, N I, O II, Si II, S I and S II as five-level atoms.
References for the atomic data used in these model atoms are
given in Lundqvist \& Fransson (1996). 
In addition, for \OII~ we  include non-thermal ionization of \OI~ to
the excited $2p^{3}~ ^{2}D^{o}$ and $2p^{3}~ ^{2}P^{o}$ levels in
\OII~(KF92).  

The following transitions are calculated in the two-level
 approximation : 
 $\CII \wl 1334$, $[\NeII] \wl 12.814 ~\mu$m, $\NaI \wl 2843$, $\MgI]
\wl 4571$. 
A-values and collision strengths are from Mendoza (1983), Fabrikant
(1974), Hayes \& Nussbaumer (1984),  Nussbaumer \& Storey (1986), and
Morton (1991).

For \SiI~ we include the $3p^{2}~^{3}P$, $3p^{2} ~^{1}D$,
 and  $3p^{2} ~^{1}S$ levels, plus  
the fine structure transitions of the $^{3}P$ ground state. 
A-values  are from Mendoza (1983), and collision
strengths for  $3p^{2}~^{3}P$ to $3p^{2} ~^{1}D$ from Pindzola,
Bhatia, \& Temkin (1977). For  the other transitions
we  estimate collision strengths from  \CI~ and \OI. 

For  $\NaI \wl 2843$,  $\MgI \wl2852$,
and $\MgII \wll2795, 2802$ we include 
non-thermal excitation with  cross sections
given in KF92.  The cross section for $\MgI] \wl4571$ is
from Fabrikant (1974).

Lines arising as a result of radiative recombination can be strong for
abundant elements, 
unless charge transfer is more efficient. An important example is
recombination to Mg I, where the strongest line is expected to be
Mg I] $\wl 4571$. No detailed calculations of the radiative recombination
cascade of this element exist. Dielectronic recombination is unimportant, although we
include it. Because most recombinations to the triplet levels should
pass through the $3p{}^3P_0$ level, we estimate the effective rate as
proportional to the statistical weight of the triplets, i.e.
$3/4$ of the total rate to the excited states, or $2.0\EE{-13}~(T/10^4
{\rm K})^{0.86} ~{\rm cm^3~s^{-1}}$.  The contribution of dielectronic recombination to the
\MgI] 4571 line is  from Nussbaumer \& Storey (1986).

\newpage

\begin{deluxetable}{lcccccc}
\tablewidth{40pc}
\tablecaption{Abundances by number in model 10H
\label{tab:10H} }
\tablehead{
\colhead{}   & \colhead{Fe--He}   & \colhead{Si--S}   &  
\colhead{O--Si--S}  & \colhead{O--C}   & \colhead{He--C}   & 
\colhead{H}}

\startdata
Mass & 0.07 & 0.30 & 1.20 & 0.60 & 2.00 & 9.00 \nl
\nl
H  & $0.00$ & $0.00$ & $0.00$ & $0.00$ & $0.00$ & $8.00\EE{-1}$ \nl
He & $5.56\EE{-1}$ & $1.48\EE{-15}$ & $5.64\EE{-12}$ & 
$6.29\EE{-10}$ & $9.75\EE{-1}$ & $2.00\EE{-1}$ \nl
C  & $2.75\EE{-4}$ & $3.81\EE{-6}$ & $1.68\EE{-2}$ & 
$1.59\EE{-1}$ & $1.82\EE{-2}$ & $2.56\EE{-5}$ \nl
N  & $0.00$ & $0.00$ & $0.00$ & $0.00$ & $6.88\EE{-6}$ & 
$1.28\EE{-4}$ \nl
O  & $3.95\EE{-7}$ & $9.38\EE{-5}$ & $8.53\EE{-1}$ & 
$8.24\EE{-1}$ & $2.03\EE{-3}$ & $1.04\EE{-4}$ \nl
Ne & $4.77\EE{-7}$ & $1.17\EE{-9}$ & $9.33\EE{-3}$ & 
$1.62\EE{-2}$ & $3.97\EE{-3}$ & $4.40\EE{-5}$ \nl
Na & $7.66\EE{-9}$ & $1.08\EE{-7}$ & $9.06\EE{-5}$ & 
$6.22\EE{-6}$ & $1.03\EE{-6}$ & $6.40\EE{-7}$ \nl
Mg & $9.38\EE{-7}$ & $1.32\EE{-5}$ & $1.11\EE{-2}$ & 
$7.62\EE{-4}$ & $1.26\EE{-4}$ & $9.60\EE{-6}$ \nl
Si & $1.10\EE{-5}$ & $6.30\EE{-1}$ & $6.46\EE{-2}$ & 
$3.79\EE{-4}$ & $1.04\EE{-4}$ & $1.36\EE{-5}$  \nl
S  & $3.42\EE{-5}$ & $2.91\EE{-1}$ & $3.88\EE{-2}$ & 
$1.90\EE{-4}$ & $5.21\EE{-5}$ & $4.48\EE{-6}$ \nl
Ar & $5.46\EE{-5}$ & $4.06\EE{-2}$ & $5.25\EE{-3}$ & 
$4.60\EE{-5}$ & $1.26\EE{-5}$ & $2.32\EE{-6}$ \nl
Ca & $1.91\EE{-4}$ & $2.58\EE{-2}$ & $5.48\EE{-4}$ & 
$2.82\EE{-5}$ & $1.90\EE{-6}$ & $7.20\EE{-7}$ \nl
Fe & $4.43\EE{-1}$ & $1.20\EE{-2}$ & $2.20\EE{-4}$ & 
$1.90\EE{-4}$ & $5.30\EE{-5}$ & $2.00\EE{-5}$ \nl
Co & -- & -- & $5.50\EE{-7}$ & $4.75\EE{-7}$ & $1.30\EE{-7}$ & $6.25\EE{-8}$ \nl
Ni &$3.56\EE{-2}$&$9.62\EE{-4}$& $1.40\EE{-5}$ & $1.20\EE{-5}$ & $3.40\EE{-6}$ & $1.60\EE{-6}$ \nl
\tablecomments{The first row gives the mass (in $\Msun$) for each
composition region. 
}
\enddata
\end{deluxetable}

\begin{deluxetable}{lcccccc}
\tablewidth{40pc}
\tablecaption{Abundances by number in model 11E1
\label{tab:11E1} }
\tablehead{
\colhead{}   & \colhead{Fe--He}   & \colhead{Si--S}   &  
\colhead{O--Ne--Mg}  & \colhead{O--C}   & \colhead{He--C}   & 
\colhead{H}}

\startdata
Mass & 0.07 & 0.10 & 1.80 & 0.10 & 2.00 & 9.00 \nl
\nl
H  & $0.00$ & $0.00$ & $0.00$ & $0.00$ & $0.00$ & $8.00\EE{-1}$ \nl
He & $7.32\EE{-1}$ & $1.73\EE{-11}$ & $8.47\EE{-7}$ & $5.90\EE{-4}$ & 
$9.81\EE{-1}$ & $2.00\EE{-1}$ \nl
C  & $6.02\EE{-4}$ & $3.64\EE{-6}$ & $8.35\EE{-3}$ & $2.71\EE{-1}$ & 
$1.41\EE{-2}$ & $2.56\EE{-5}$ \nl
N  & $0.00$ & $0.00$ & $0.00$ & $0.00$& $4.76\EE{-6}$ & $1.28\EE{-4}$ \nl
O  & $8.70\EE{-7}$ & $1.92\EE{-5}$ & $7.83\EE{-1}$ & $7.15\EE{-1}$ & 
$1.93\EE{-3}$ & $1.04\EE{-4}$ \nl
Ne & $3.81\EE{-7}$ & $3.31\EE{-10}$ & $1.14\EE{-1}$ & $1.06\EE{-2}$ & 
$2.63\EE{-3}$ & $4.40\EE{-5}$ \nl
Na & $1.28\EE{-10}$ & $0.00$ & $5.18\EE{-4}$ & $5.90\EE{-6}$ & 
$1.70\EE{-6}$ & $6.40\EE{-7}$ \nl
Mg & $1.02\EE{-6}$ & $1.44\EE{-5}$ & $7.12\EE{-2}$ & $2.72\EE{-3}$ & 
$2.40\EE{-5}$ & $9.60\EE{-6}$ \nl
Si & $1.47\EE{-5}$ & $5.87\EE{-1}$ & $1.64\EE{-2}$ & $1.30\EE{-4}$ & 
$3.60\EE{-5}$ & $1.36\EE{-5}$ \nl
S  & $7.36\EE{-5}$ & $2.76\EE{-1}$ & $2.10\EE{-4}$ & $4.20\EE{-5}$ & 
$1.20\EE{-5}$ & $4.48\EE{-6}$ \nl
Ar & $1.86\EE{-4}$ & $4.04\EE{-2}$ & $2.40\EE{-5}$ & $2.10\EE{-5}$ & 
$6.00\EE{-6}$ & $2.32\EE{-6}$ \nl
Ca & $6.56\EE{-4}$ & $2.64\EE{-2}$ & $7.70\EE{-6}$ & $6.70\EE{-6}$ & 
$1.90\EE{-6}$ & $7.20\EE{-7}$ \nl
Fe & $2.39\EE{-1}$ & $6.94\EE{-2}$ & $2.10\EE{-4}$ & $1.80\EE{-4}$ & 
$5.20\EE{-5}$ & $2.00\EE{-5}$ \nl
Co & -- & -- & $5.25\EE{-7}$ & $4.50\EE{-7}$ & $1.30\EE{-7}$ & $6.25\EE{-8}$ \nl
Ni &$2.02\EE{-2}$&$5.69\EE{-3}$ & $1.40\EE{-5}$ & $1.20\EE{-5}$ & $3.40\EE{-6}$ & $1.60\EE{-6}$ \nl
\tablecomments{The first row gives the mass (in $\Msun$) for each
composition region. 
}
\enddata
\end{deluxetable}

\begin{deluxetable}{lc}
\tablewidth{20pc}
\tablecaption{Optical depths to the gamma-rays at 500 days. \label{tab:gamma} }
\tablehead{
\colhead{Zone}   & \colhead{$\Delta \tau_{\gamma}$ }
}
\startdata
Fe -- He & 0.015 \nl
Si -- S  & 0.061 \nl
O -- Si -- S & 0.225 \nl
O -- C & 0.103 \nl 
He (core) & 0.090 \nl
He (envelope) & 0.030 \nl
H (core) & 0.676 \nl
H (envelope) & 0.302 \nl
& \nl
Total & 1.502 \nl
\enddata
\end{deluxetable}

\begin{deluxetable}{lclc}
\tablewidth{40pc}
\tablecaption{Charge Transfer Reactions
\label{tab:ct} }
\tablehead{
\colhead{Reaction}   & \colhead{Reference}   & \colhead{Reaction}   &  
\colhead{Reference}}

\startdata
H I + He II $\rightarrow$  H II + He I  & 1 &
H I + He III $\rightarrow$ H II + He II & 1 \nl
H I + C II  $\rightarrow$ H II + C I    & 1 &
H I + C III $\rightarrow$ H II + C II   & 1 \nl
H I + N II  $\rightarrow$ H II + N I    & 1 &
H I + O II  $\rightarrow$ H II + O I    & 1 \nl
H I + O III $\rightarrow$ H II + O II   & 1 &
H I + Mg III $\rightarrow$ H II + Mg II & 1 \nl
H I + Si III $\rightarrow$ H II + Si II & 1 &
H I + S II  $\rightarrow$ H II + S I    & 1 \nl
H I + S III $\rightarrow$ H II + S II   & 1 &
H I + Fe III $\rightarrow$ H II + Fe II & 1 \nl
H I + Fe IV $\rightarrow$ H II + Fe III & 1 &
H I + Fe V $\rightarrow$ H II + Fe IV   & 1 \nl
H II + He I $\rightarrow$ H I + He II   & 4 &
H II + C I  $\rightarrow$ H I + C II    & 1 \nl
H II + N I  $\rightarrow$ H I + N II    & 1 &
H II + O I  $\rightarrow$ H I   + O II  & 1 \nl 
H II + Na I $\rightarrow$ H I + Na II   & 5 &
H II + Mg I $\rightarrow$ H I + Mg II   & 1 \nl
H II + Mg II $\rightarrow$ H I + Mg III & 1 &
H II + Si I $\rightarrow$ H I + Si II   & 2 \nl
H II + Si II $\rightarrow$ H I + Si III & 1 &
H II + S I $\rightarrow$ H I + S II     & 1 \nl
H II + Ca I $\rightarrow$ H I + Ca II   & 9 &
H II + Fe I $\rightarrow$ H I + Fe II   & 3 \nl
H II + Fe II $\rightarrow$ H I + Fe III & 1 &
He I + C III $\rightarrow$ He II + C II & 2 \nl
He I + O III $\rightarrow$ He II + O II & 2 &
He II + C I $\rightarrow$ He I + C II   & 6 \nl
He II + C II $\rightarrow$ C III + He I & 2 &
He II + Si I $\rightarrow$ He I + Si II & 3 \nl
He II + Si II $\rightarrow$ Si III + He I & 2 &
He II + S II  $\rightarrow$ S III + He I  & 2 \nl
C II + Na I $\rightarrow$ Na II + C I   & 3 &
C II + Mg I $\rightarrow$ Mg II + C I   & 3 \nl
C II + Si I $\rightarrow$ Si II + C I   & 3 &
C II + S I $\rightarrow$ S II + C I     & 3 \nl
C II + Fe I $\rightarrow$ Fe II + C I   & 3 &
N II + Mg I $\rightarrow$ N I + Mg II   & 3 \nl
N II + Ca I $\rightarrow$ N I + Ca II   & 7 &
N II + Fe I $\rightarrow$ N I + Fe II   & 3 \nl
O II + Ca I $\rightarrow$ O I + Ca II   & 8 &
O II + Fe I $\rightarrow$ O I + Fe II   & 3 \nl
Na I + Mg II $\rightarrow$ Na II + Mg I & 3 &
Na I + Si II $\rightarrow$ Na II + Si I & 3 \nl
Na I + S II $\rightarrow$ Na II + S I   & 3 &
Na I + Fe II $\rightarrow$ Na II + Fe I & 3 \nl
Mg I + Si II $\rightarrow$ Mg II + Si I & 3 &
Mg I + S II $\rightarrow$ Mg II + S I   & 3 \nl
Si I + S II $\rightarrow$ Si II + S I   & 3 &
Si II + Fe I $\rightarrow$ Si I + Fe II & 3 \nl
S II + Fe I $\rightarrow$ S I + Fe II   & 3 &
\tablerefs{(1) Kingdon \& Ferland 1996; (2) Arnaud \& Rothenflug 1985;
(3) Prasad \& Huntress 1980;
(4) Kimura, Lane, Dalgarno, \& Dixson 1993; (5) Croft \& Dickinson 1996;
(6) Kimura \etal 1993; (7) Swartz 1994; (8) Rutherford \& Vroom (1972); 
(9) Estimate; the same rate as for Ca I + O II}
\enddata
\end{deluxetable}

\newpage

\end{document}